# Van der Waals superconducting electronics: materials, devices and circuit integration

A. Di Bernardo[1,2], E. Scheer[1]

1. University of Konstanz, Universitätsstraße 10, 78464 Konstanz, Germany
2. University of Salerno, Department of Physics 'E. R. Caianiello', 84084 Fisciano (SA), Italy

Correspondence to: adibernardo@unisa.it or angelo.dibernardo@uni-konstanz.de

**Abstract**

Van der Waals (vdW) superconductors — atomically thin crystalline materials that can be stacked into more complex heterostructures — have opened a promising avenue for superconducting electronics thanks to their properties that are otherwise difficult to obtain in other superconducting materials. These include strong resilience to high in-plane fields, electrostatic tuneability, and non-reciprocal transport rooted in inversion-symmetry breaking and strong spin–orbit coupling. In addition to highlighting the importance of these properties for superconducting electronics, this review gives an overview over the physical mechanisms that govern and influence superconductivity in vdW materials including Ising pairing, band inversion, and proximity effects at superconductor/ferromagnet interfaces that do not have an equivalent in thin-film systems. This overview then sets the basis to survey the wide range of functionalities enabled by superconducting vdW devices including gate-controlled devices, superconducting diodes, and circuit elements for readout and control of quantum bits. The review concludes with a forward look at wafer-scale growth and deterministic assembly of vdW devices, highlighting concrete pathways that can enable the transition from vdW device prototypes to deployable components for cryogenic electronics and quantum technologies.

# 1. Introduction

More than a century after its discovery and several decades after its first microscopic theory, superconductivity still reveals unknown fascinating phenomena when realised in new materials. Van der Waals (vdW) superconductors (Ss) and their heterostructures have rapidly emerged as versatile platforms for exploring new regimes of quantum matter. Thanks to their atomically flat interfaces, absence of dangling bonds and the ease in assembling them into layered stacks, vdW Ss enable device concepts that are difficult to realise with three-dimensional (3D) S thin films. In particular, vdW Ss provide unprecedented opportunities to study and exploit phenomena like topological superconductivity (TS), unconventional pairing mechanisms, proximity effects (PEs), and gate-control of superconductivity in regimes that are not accessible with 3D Ss, and which can prove key for the development of the next-generation of superconducting electronics.

Transition metal dichalcogenides (TMDs with general formula $MX_2$, where M = Nb, Ta, Mo, W etc. is the transition metal and X = S, Se, Te) are the prototypical example of vdW Ss. Several TMDs (e.g., $NbSe_2$, $NbS_2$, $TaS_2$) are intrinsically superconducting down to the few-layer limit, while others (e.g., $WTe_2$, $MoTe_2$ in specific phases) host band inversions associated to non-trivial topology which makes them interesting platforms to study the interplay between topology and superconductivity [1], for example, in proximity effect (PE) studies (see Box 1). Throughout this review, we will consider both devices entirely made of vdW materials and hybrid architectures of 3D/vdW materials. In the latter type of devices, a vdW material provides the active region or weak link of the device such as, for example, the channel of a Josephson junction (JJ) or the sensing element of a superconducting quantum interference device (SQUID), whilst a 3D S is used to induce superconductivity in it via the PE.

The timeliness of this review is underscored by the fast pace of recent developments in the fields of superconducting electronics and quantum computing based on vdW Ss. Over the past few years, different groups have reported evidence consistent with TS in magnetic–superconducting vdW hybrids and at engineered edges in vdW heterostructures [2-4]. In parallel, studies of Ising pairing in vdW Ss have revealed in-plane critical fields far exceeding the Pauli limit in few-layer vdW Ss and related systems [5-9], while long-ranged spin-triplet supercurrents – central to superconducting spintronics – have recently been demonstrated also for vdW superconductor/ferromagnet (S/F) hybrids [10,11].

Concurrently, a suite of vdW-enabled superconducting devices has matured: devices based on modulation of the superconducting state with a gate voltage ($V_G$) [12-14], SQUIDs based on twisted vdW systems or with novel vdW-enabled geometries [15,16], vdW-based bolometers and quantum-limited parametric amplifiers [17-19], and non-reciprocal diode elements without external magnetic fields in vdW JJs [20]. These results highlight the promise of vdW Ss for information processing, sensing and low-power logics and motivate an assessment of current capabilities and outstanding challenges related to the scalability and integration of vdW-based devices into existing superconducting digital and quantum computing architectures.

The purpose of this review is therefore twofold: first, to summarise the experimental and theoretical advances that define the current landscape of vdW superconductivity and its device applications; and second, to identify the key open questions and materials-science bottlenecks that must be addressed to translate recent device demonstrators into actual circuits. Compared to ref. [21], which mainly surveys vdW-based JJs and devices, in this review we discuss the physical properties of vdW Ss and phenomena that can be potentially realised with them, and also survey a wider landscape of devices for vdW superconducting electronics including gate-tuneable elements and diodes as well as superconducting devices based on PEs in vdW S/F hybrids.

The review starts with an overview of recent efforts made to realise TS in vdW heterostructures and reporting the status of experimental evidence for Majorana modes in vdW systems (section 2). We then examine the mechanisms of unconventional pairing and the remarkable resilience of vdW Ss to the application of high in-plane magnetic fields (section 3). Subsequent sections review the rapidly growing field of vdW S/F and their prospects for superspintronics (section 4), as well as vdW Josephson devices for quantum sensing and microwave applications (section 5). We further consider vdW hybrid systems incorporating molecules (section 6), non-reciprocal charge transport in vdW superconducting diodes (section 7), and the tuning of superconductivity in three-terminal (gated) vdW devices (section 8). Finally, we address materials-engineering aspects including large-scale fabrication (section 9) and integration of vdW superconducting elements into superconducting quantum-circuit architectures (section 10) and conclude with an outlook on future research directions (section 11).

## 2. Topological superconductivity enabled by vdW heterostructures

Superconducting vdW heterostructures offer exciting potential for the realisation of TS [22], which is in turn promising for fault-tolerant quantum computing [23]. In the context of TS, a key advantage offered by vdW heterostructures over equivalent systems based on thin films from 3D materials stems from the fact that the edges of vdW stacks, where TS should manifest, are easily accessible. The same edges can be engineered to be not buried, if the lateral arrangement of the vdW flakes (within a vdW stack) is properly designed. The latter aspects also facilitate the study and manipulation of topological edge states in vdW systems under electrical or optical stimuli.

Kezilebieke et al. [2] recently reported spectroscopic evidence consistent with the emergence of TS in a monolayer of the vdW ferromagnetic insulator (FI) $CrBr_3$ coupled to the vdW S $NbSe_2$. By performing low-temperature scanning tunnelling microscopy (low-$T$ STM) measurements, the authors resolved zero-bias conductance peaks (ZBCPs) in the differential conductance, $dI/dV$, of this vdW system, which they argued to be signature for Majorana topological edge states [23]. For an introduction to Majorana states and to the spectral features associated with their emergence in the $dI/dV$ of a superconducting system, see Box 1. One of the main reasons for the claim on the observation of Majorana modes in ref. [2] is that the ZBCPs only appeared at the $CrBr_3$ island edge, whilst they disappeared as the STM tip was moved further away from the edges or when superconductivity in $NbSe_2$ was suppressed by the application of a sufficiently high external magnetic field ($B_{ext}$).

Similar results were also reported in a different study for a nanosheet of the antiferromagnetic insulator (AFI) $CrI_2$ coupled to $NbSe_2$ (S) [4]. The $dI/dV$ spectra of this AFI/S vdW heterostructure showed in-gap states in the superconducting density of states (DoS), which got stronger at the edge between the two vdW materials and disappeared ~ 9 nm away from the edge. Nonetheless, unlike for the CrBr$_3$/NbSe$_2$ (FI/S) system studied in ref. [2], no ZBCPs were observed in ref. [4]; instead, the CrI$_2$/NbSe$_2$ (AFI/S) vdW hybrid displayed a continuous sub-gap spectral feature dispersing across the entire superconducting gap ($\Delta$) and with only a faint intensity at zero bias. Based on a phenomenological model assuming a dominant *p*-wave component in the superconducting order parameter, the authors suggested that a topological phase with Chern number $C = 2$ can be stabilised in the CrI$_2$/NbSe$_2$ hybrid. The even parity of $C$ is justified by the antiferromagnetic order of CrI$_2$, which forces the number of Fermi surfaces to be even. The model also predicts two chiral edge modes, which can lead to the increase in the sub-gap differential conductance observed experimentally [4].

In another study [24] also done on islands of a vdW AFI (CrBr$_3$ in this case) on the vdW S NbSe$_2$, the authors observed two types of edge states in the d$I$/d$V$ spectra measured by low-$T$ STM. One of the edge states preserves particle-hole symmetry, whilst the other produces a ZBCP within the gap. These features were discretely distributed at the edges of the CrBr$_3$/NbSe$_2$ islands (figures 1(A)-(C)). The authors therefore concluded that the observed edge states were not Majorana states which should be instead topologically protected and insensitive to disorder. The authors also argued that the lack of evidence for changes in the DoS at the CrBr$_3$/NbSe$_2$ edges in earlier studies like ref. [2] was possibly due to the poorer spatial resolution of the STM used compared to their setup. By contrast, the discontinuous variation in the d$I$/d$V$ observed in ref. [24] was explained as the result of lattice reconstructions at the CrB$_3$/NbSe$_2$ step edge, and the ZBCPs were interpreted as a signature of Yu-Shiba-Rusinov (YSR) states [25-27] (see Box 1 for further details on YSR states).

Another vdW material studied for the realisation of TS is the semimetal WTe$_2$. Whilst bulk semimetallic states dominate electronic transport in the normal state of WTe$_2$, topological gapless (hinge) states are predicted to dominate electronic transport once superconductivity is induced in WTe$_2$. Although, to the best of our knowledge, no JJs with vdW Ss and a WTe$_2$ weak link have been studied to date, several groups have studied JJs with 3D S electrodes (usually Pd or Nb) and a WTe$_2$ weak link [3,28-30]. These studies report an unusual current-phase relation (CPR) – namely, the dependence of the critical current $I_c$ on the phase difference between the two Ss – with a $4\pi$-periodicity which has been interpreted as a signature for TS. In a JJ with a topological weak link, perfect Andreev reflections (see Box 1) should in fact occur at each S/weak link interface due to spin-momentum locking in the hinge states which prevents normal electron reflection. For a long JJs, this mechanism leads to a CPR with a sawtooth shape and $4\pi$-periodicity, and the supercurrent must be carried by Andreev bound states (ABS) with opposite parity [31]; see figures 1(F) and (H).

Nonetheless, TS is not the only possibility to explain such CPR in a JJ with a WTe$_2$ weak link. In ref. [30] the authors argued that the same CPR can also stem from additional junctions forming at the interface between the 3D Ss and the TI WTe$_2$ (figures 1(F)-(I)). This explanation was proposed in particular for JJs where Pd was used as the 3D S, where additional superconducting PdTe$_x$ phases can form at the Pd/WTe$_2$ interfaces due to the Pd interdiffusion into the WTe$_2$ weak link [30].

To summarise, despite the numerous reports claiming TS in vdW systems, by the time of writing this review, an unambiguous piece of evidence for the formation of topological edge states in vdW systems still seems to be lacking. Better controlled studies are therefore needed to fully establish the potential that vdW systems hold for the realisation and manipulation of vdW-based TS.

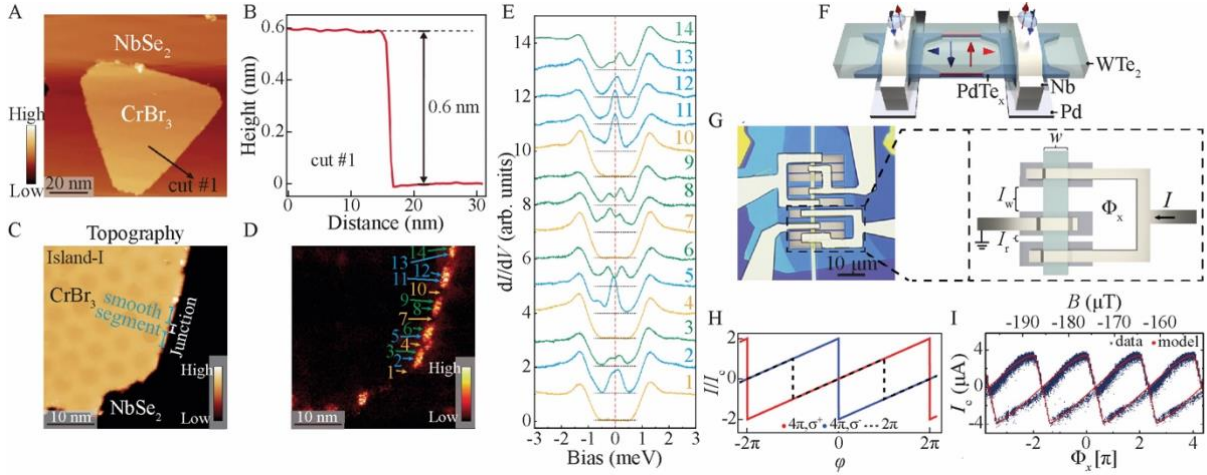

**Figure 1.** Evidence for edge states in vdW heterostructures. (A)-(B) CrBr$_3$/NbSe$_2$ heterostructure probed by low-temperature STM: (A) topographic image of a monolayer of CrBr$_3$ island on NbSe$_2$ with (B) line profile taken along cut#1 in (A). (C) Topographic image of a second CrBr$_3$ island with (E) corresponding differential conductance d$I$/d$V$ traces taken at the numbered position in (D) highlighting spectral features at the edge of the CrBr$_3$/NbSe$_2$ vdW interface. (F)-(I) VdW WTe$_2$ topological Josephson junction (JJ) and asymmetric SQUID readout: (F) JJ schematic with WTe$_2$ weak link bridging Pd contacts; (G) device optical image (left) and SQUID layout (right; dashed box); (H) theoretical current versus phase relation normalised to the critical current, $I(\varphi)/I_c$, for a topological JJ; (I) experimental $I_c(\varphi)$ of the SQUID in (G) embedding the WTe$_2$-based JJ. Panels (A)-(E) adapted from ref. [24], panels (F)-(I) from ref. [30].

## 3. Unconventional pairing and resilience to magnetic fields in vdW superconductors

A notable property of vdW Ss is their high in-plane upper critical field ($B_{c2//}$), which can be exploited to develop superconducting devices that are robust in an applied $B_{ext}$ and more resilient to magnetic noise compared to similar devices made from 3D S thin films. Both non-centrosymmetric and symmetric vdW Ss have been reported with $B_{c2//}$ greatly exceeding the so-called Pauli field ($B_p$) – this is the magnetic field that is required to suppress superconductivity via spin-flipping of electrons within conventional spin-singlet Cooper pairs (CPs) [32,33].

The enhancement in $B_{c2//}$ in vdW Ss compared to 3D Ss is attributed to several factors but mostly related to their electronic band structure. In non-centrosymmetric vdW Ss, spin-orbit coupling (SOC) lifts the spin degeneracy of the electronic bands which leads to Zeeman-protected superconductivity and, in turn, to an increase in $B_{c2//}$. A SOC-driven enhancement in $B_{c2//}$ has been reported for vdW Ss like NbSe$_2$ [5,34] and TaS$_2$ [35], as well as for other materials like MoS$_2$ and WS$_2$ that are normally semiconducting but can be driven superconducting under the application of an electric field, e.g., with ionic liquid gating [6,7,36].

**Box 1. Theoretical concepts useful for the review.**

Topological superconductivity and Majorana modes. The expression 'topological superconductivity' refers to a superconducting phase that is characterised by non-trivial topology leading to protected boundary excitations [22]. In one- or two-dimensional systems, these protected excitations are called Majorana zero modes (MZMs). MZMs are self-conjugated particles the existence of which stems from a combination of superconducting pairing, spin-orbit interaction and Zeeman coupling of the spins [37].
In a topological superconductor, the bulk gap protects MZMs against perturbations. This protection, in combination with the non-Abelian exchange statistics and non-local nature of MZMs can provide a route to fault-tolerant computing [38,39]. Two widely-discussed signatures of MZMs are a zero-bias conductance peak (ZBCP) in the superconducting density of states (DoS) with height approaching $2e^2/h$, which can be measured by tunnelling spectroscopy ($h$ being the Planck's constant) [40], and a Josephson effect with $4\pi$ periodicity measurable in a Josephson junction (JJ) with a topological weak link [41].

Scanning tunnelling spectroscopy and sub-gap features. In scanning tunnelling spectroscopy, the differential conductance, d$I$/d$V$ (**r**, $V_b$), measured at a position identified by the space vector **r** and bias voltage $V_b$ – applied between a tip and a superconducting sample – is proportional to the local DoS of the superconductor (S) [42]. In addition to giving information about the size of the superconducting gap, the DoS can also contain sub-gap features that provide evidence for the emergence of different states in a superconducting system like MZMs, Andreev bound states (ABSs) and Yu-Shiba-Rusinov (YSR) states.
YSR states are bound to a magnetic impurity on the surface of a S, are spin-polarised and can be measured only locally around the impurity [43-45]. The sub-gap features associated with YSR states appear on both voltage polarities but are usually asymmetric in amplitude and split linearly in an applied magnetic field ($B_{ext}$), if the impurity is screened [46,47].
ABSs emerge as result of Andreev reflections (see below) occurring at the interface between a S and a normal metal (N) or semiconductor (SM). Sub-gap features due to ABSs in the superconducting DoS can move under the application of a $B_{ext}$ due to Zeeman splitting and merge leading to a ZBCP mimicking that of a MZM [48].
MZMs have a non-local nature and, unlike YSR or ABSs, appear at both ends of a topological segment. MZMs are not very sensitive to small changes in the local chemical potential (albeit not immune [48]) and give rise to a ZBCP that remains pinned around $V_b = 0$ for a certain $B_{ext}$ window [49,50]. The orientation of the applied $B_{ext}$ also matters because a $B_{ext}$ parallel to the spin-orbit axis maximises the topological gap [49]. Correlated end-to-end conductance or perturb-one-end/measure-the-other protocols can help differentiate MZMs from ABS or YSR states [48,51].

Superconducting proximity effect, Andreev reflections and spin-triplet superconductivity. The superconducting proximity effect originates from Andreev reflections taking place at an interface between a S and another material X, where X can be a SM, a N or a ferromagnet (F). In an Andreev reflection, an electron incident from the material X into S is retroreflected as a hole with an opposite spin, while a CP gets transferred into S [52,53]. The probability of this process is set by the interface transparency and the normal-conductive-property mismatch of the two materials. In diffusive systems these are described by the conductivities and the diffusion constants, in ballistic systems by the Fermi velocities. Multiple Andreev reflections in S/N/S JJs with a short N weak link generate ABSs that have energies dispersed with phase and produce a subharmonic gap structure at energies $E_n = 2\Delta/n$, where $\Delta$ is the superconducting gap amplitude and $n$ is an integer [54]. Inside a diffusive N proximitized to a S, the anomalous pair amplitude decays over the coherence length $\xi_N = \sqrt{\hbar D/2\pi k_B T}$ ($D$ being the diffusion constant in N, $\hbar$ the reduced Planck's constant, $k_B$ the Boltzmann's constant, and $T$ the temperature) and opens a minigap of amplitude $E_g \lesssim E_{Th} = \hbar D/L^2$ ($L$ being the weak link length) in the local DOS, where $E_{Th}$ is the Thouless energy [55,56]. For a ballistic $N$, $\xi_N = \hbar v_F/2\pi k_B T$, where $v_F$ is the Fermi velocity [57]. At the same time, the inverse proximity effect at the S/N interface suppresses $\Delta$ on the S side [57].
At S/F interfaces, the exchange field $h_{ex}$ of the F suppresses conventional (spin-singlet) Andreev processes and causes oscillatory and rapidly-decaying spin-singlet and spin-triplet correlations, which results in a change in the coupling between the two Ss in S/F/S JJ from 0 to $\pi$, as the thickness of F is increased [58,59]. A magnetically inhomogeneous F at a S/F interface can convert spin-singlet CPs into odd-frequency equal-spin (spin-triplet) CPs that are insensitive to $h_{ex}$ and can hence propagate inside F over much longer distances than spin singlets [60] — this effect is central to superspintronics.

Monolayer NbSe$_2$ is probably the most remarkable example of vdW S with $B_{c2//}$ greatly exceeding $B_P$. The increase in $B_{c2//}$ above $B_P$ for monolayer NbSe$_2$ has been attributed to the stabilisation of Ising-type superconductivity, which involves the formation of superconducting pairing correlations between electrons with antiparallel spins locked perpendicular to the surface of the NbSe$_2$ in the two-dimensional (2D) limit. For 2D (monolayer) NbSe$_2$, the in-plane mirror symmetry is broken by the hexagonal symmetry of the lattice combined with a unit cell structure consisting of one layer of Nb sandwiched between two Se layers. The broken mirror symmetry together with the large SOC of Nb induces spin splitting of the electronic states at finite momentum $k$. This spin splitting can be seen as the result of an effective magnetic field $B_{so}$ that must be oriented perpendicular to both the electron's motion and to the crystal field. Since in a monolayer, both the electron's motion and the crystal field lie within the material's plane (the latter due to the broken mirror symmetry), $B_{so}$ must be oriented out-of-plane. The out-of-plane $B_{so}$ locks the electrons' spins out-of-plane, with opposite-momentum electrons having spins pointing along different directions due to spin-valley coupling [61,62]. CPs form between electrons with out-of-plane spins at the K and K' point of the Brillouin zone, where the spin splitting is greater than at the Γ point [5], which leads to the so-called Ising pairing (figure 2(A)).

In a 2D S, $B_{c2//}$ should be already much larger than the $B_{ext}$ required to destroy superconductivity due to orbital depairing because orbital effects are absent when $B_{ext}$ is applied in-plane. As a result, $B_{c2//}$ at $T = 0$, $B_{c2//}(0)$, should in principle match $B_P$ at $T = 0$, $B_P(0)$ – this is equal to $(1.76\, k_B T_c)/\sqrt{2}\mu_B \approx 1.84\, T_c$ according to Bardeen Cooper Schrieffer (BCS) Ss [63], where $k_B$ is the Boltzmann constant and $\mu_B$ the Bohr magneton. Nonetheless, for monolayer NbSe$_2$, the measured $B_{c2//}(0)$ is about six times larger than $B_P(0)$, as result of Ising pairing [5] (see figures 2(B) and (C)).

An enhancement of $B_{c2//}$ above $B_P$ has not only been measured in non-centrosymmetric vdW Ss like NbSe$_2$ but also for centrosymmetric Ss. Centrosymmetric Ss with $B_{c2//} >> B_P$ also include 3D Ss (i.e., non-vdW) like PdTe$_2$ [9] and epitaxially strained α-Sn (or stanene) [8] grown in the form of thin films. Here, the enhancement in $B_{c2//}$ stems from spin-orbit locking, which causes spin bands at the Γ point to experience opposite effective out-of-plane magnetic fields. This spin-splitting occurs without inversion symmetry breaking. The CPs form between electrons moving on orbits around the Γ point in spin-split bands that pair up with their spins aligned out-of-plane. Given the out-of-plane direction of the paired electrons' spins, in analogy with Ising superconductivity, this type of superconductivity has been named type-II Ising superconductivity [8].

Similar to their 3D counterpart, centrosymmetric vdW Ss can also show a $B_{c2//}$ above $B_p$. However, centrosymmetric vdW Ss have been investigated in more detail only recently, also due to the increasing interest in TS, which some of these Ss are predicted to host. Centrosymmetric vdW Ss like 1T'-WTe$_2$ [12,13,66-68] and 2M-WS$_2$ [65,69], for example, feature topological band inversion. When the 2D limit is reached for these vdW Ss, conventional SOC terms involving only spin and momentum should be forbidden by inversion symmetry. Nevertheless, spin, momentum, and parity can still couple near those points in momentum space where bands with opposite parities invert and open a topological gap [65]. Such coupling between spin, momentum and parity of the electronic states is known as spin-orbit-parity coupling (SOPC) [64], as shown in figure 2(D). In ref. [65], the authors showed that few-layer-thick 2M-WS$_2$ exhibits $B_{c2//}(0) \sim 30.51$ T (figures 2(E)-(H)), which is larger than $B_P(0)$. In addition, $B_{c2//}(0)$ displays a twofold anisotropy, when measured as a function of the angle between an applied in-plane $B_{ext}$ and the $a$-axis of the crystal. According to the authors, the absence of inversion symmetry breaking and spin-orbit locking rules out both type-I and type-II Ising superconductivity as mechanisms responsible for the increase in $B_{c2//}(0)$. Instead, the enhancement in $B_{c2//}(0)$ is ascribed to the SOPC [65].

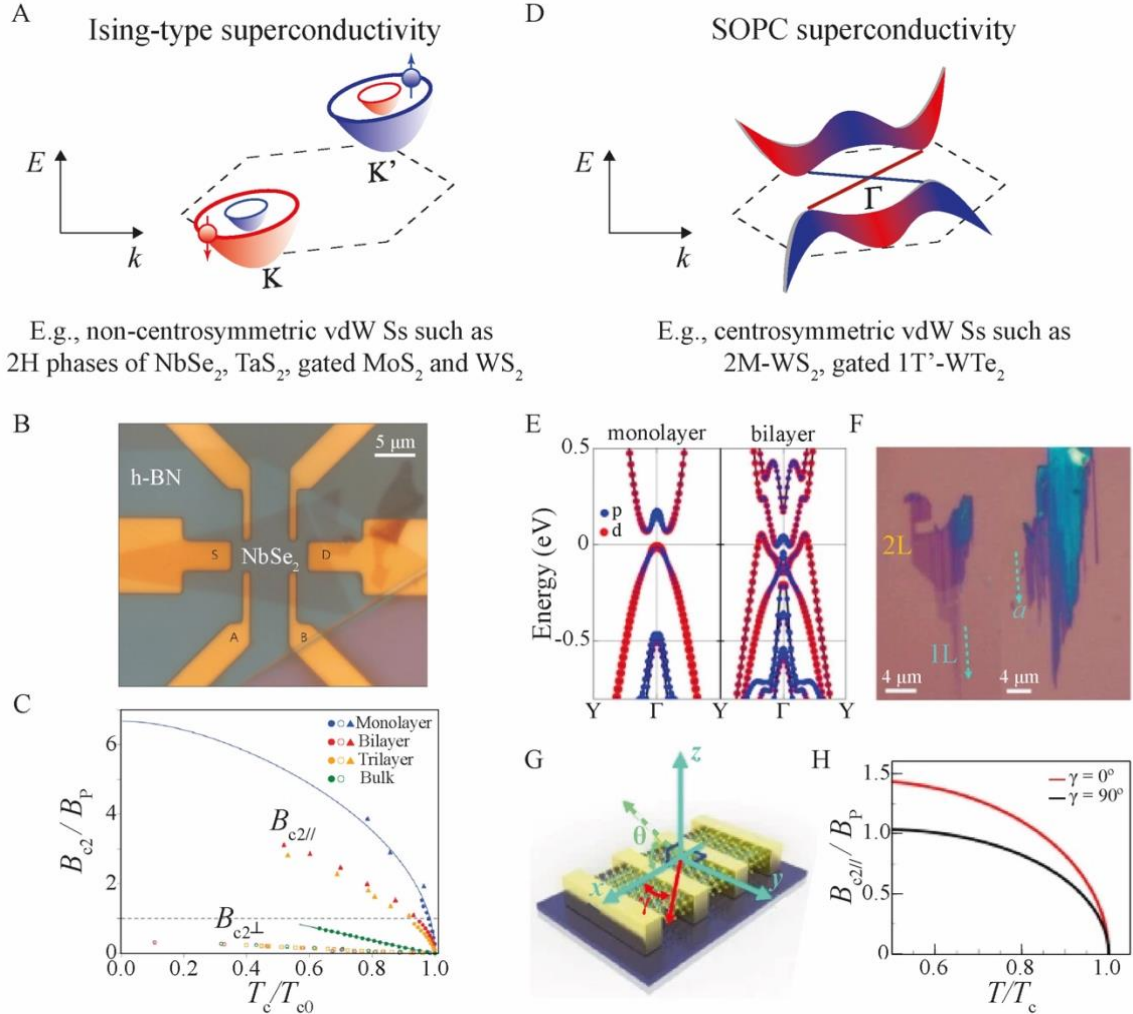

**Figure 2.** Ising and spin-orbit-parity coupled (SOPC) superconductivity in vdW superconductors. (A) Type-I Ising pairing: energy $E$ versus momentum $k$ diagram showing pairing correlations forming between electrons belonging to opposite-spin valleys at K and K'. (B) Optical image of NbSe$_2$ bilayer and (C) upper critical field $B_{c2}$ versus superconducting critical temperature $T_c$ for monolayer (blue data), bilayer (red), trilayer (orange) and bulk (green) NbSe$_2$. In-plane $B_{c2}$, $B_{c2//}$, (filled symbols) and out-of-plane $B_{c2}$, $B_{c2\perp}$, (hollow symbols) in (C) are normalised to the Pauli field $B_P$ whilst $T_c$ is normalised to its value without any applied magnetic field $T_{c0}$. (D) SOPC superconductivity: schematic with even-parity (red) and odd-parity (blue) bands. (E) Calculated density of states for W $d$-orbitals (red) and S $p$-orbitals (blue) in mono- and bilayer 2M-WS$_2$. (F)-(H) 2M-WS$_2$ flakes: (F) optical image; (G) device geometry for measurement of angular dependence of $B_c$; (H) $B_{c2//}$ versus $T_c/T_{c0}$ for magnetic field applied along the $a$-axis ($\gamma = 0°$; red) and $b$-axis ($\gamma = 90°$; black). Panel (A) adapted from [8], panels (B)-(C) from [5], panel (D) from [64], panels (E)-(H) from [65].

## 4. VdW Superconductor/ferromagnet hybrids for superspintronics

Superconducting vdW heterostructures also represent an interesting platform to study the interplay between unconventional superconductivity (e.g., Ising superconductivity) and magnetism, and thereby to discover novel physical phenomena as well as to engineer and test novel superconducting spintronic (superspintronic) devices based on PEs at S/F vdW interfaces. In superspintronics, Ss and Fs are combined to generate CPs with parallel-aligned spins (i.e., fully polarised spin-triplet CPs; see also Box 1), which can be used to do spintronics in the superconducting state where energy dissipation by Joule heating is virtually absent.

Compared to conventional 3D S/F thin film heterostructures, the intrinsic crystallinity of vdW flakes and their reduced lateral size offer several advantages for device fabrication including the possibility to avoid high-temperature growth processes – which are usually required for the synthesis of 3D single-crystalline Ss – and to reduce the patterning steps needed to define the device geometry. The small amount of magnetic volume present in a few-layer-thick vdW F flake can also help address another major superspintronic challenge: demonstrating that a spin-triplet supercurrent generated at a S/F interface can be used to switch the magnetisation of the F layer by exerting spin-transfer torque onto it [70,71].

A key prerequisite for the realisation of superconducting devices based on S/F vdW systems with a metallic F is the need to form a clean S/F vdW interface to ensure a strong PE. Achieving an S/F interface with good transparency and perfect engagement between the S and F vdW flakes, however, can be tricky and very sensitive to the specific fabrication process followed. In ref. [72] the authors studied how the fabrication process affects the PE in the vdW ferromagnetic metal $Fe_3GeTe_2$ (FGT) coupled to the vdW S $NbSe_2$. High-resolution transmission electron microscopy revealed the formation of a polymeric layer at the $NbSe_2$/FGT interface made of residues of the polydimethylsiloxane (PDMS) used in the dry-transfer assembly of the S/F vdW stack. Despite the presence of these PDMS residues, the authors could still observe a superconducting state induced in FGT, when the thickness of FGT was reduced below 4 nm. To selectively measure the resistance of the FGT, the vdW stack was assembled with $NbSe_2$ on top and with FGT at the bottom and in contact with pre-patterned (bottom) Au electrodes. In devices with thicker FGT layer, the resistance increased as the sample's temperature was decreased below the $T_c$ of $NbSe_2$, meaning that the FGT flake was not made fully superconducting by the PE.

In the same study [72], Hu and co-workers also reported the differential resistance, $dV/dI$, measured for $NbSe_2$/FGT vdW samples with a 4-nm-thick FGT layer. They identified two peak features with amplitudes of 1.22 and 0.66 meV which, based on their dependence on $T$ and $B_{ext}$, could be attributed to multiple Andreev reflections occurring at the S/F vdW interface. In addition, the authors observed a subgap of 48 μeV, consistent in size with the proximity-induced minigap reported in other studies for the $NbSe_2$/$WTe_2$ vdW system [73]. This subgap was interpreted as the minigap induced in FGT by the PE with $NbSe_2$. Using the polymeric layer as an insulating spacer (I) and the fully-proximitized 4-nm-thick FGT as a S' layer, the authors also succeeded in measuring a modulation of the $I_c$ of the $NbSe_2$/spacer/FGT (S/I/S') stack under an applied out-of-plane $B_{ext}$. This result suggests that, although polymeric residues can suppress the PE, they can also be used as a weak link for CP tunnelling when they are not too thick.

In analogy with 3D superspintronics, also for S/F vdW stacks, changes in the magnetisation of the vdW F can be used to modulate the $T_c$ of the vdW S coupled to it. Significant variations in $T_c$ (or in the resistance $R$ measured at a specific temperature $T$ across the superconducting transition) set by specific states of the vdW

F's magnetisation can be exploited for the realisation of logic devices operating at cryogenic temperatures. These logic devices are appealing when the variation in $T_c$ values (or $R$) associated with the different logic states are much larger than those that can be achieved due to thermal fluctuations in the cryostat, and also when the $T_c$ (or $R$) values are well-reproduced once the F's magnetisation is brought back into a specific configuration by the applied $B_{ext}$.

Jo et al. measured the magnetoresistance (MR) of a CrSBr/NbSe$_2$ (FI/S) vdW device at a fixed $T$ (~ 5 K) across its superconducting transition [74]. In their work, the MR is defined as MR = ($R_{max}(B_{ext})$-$R_{min}(B_{ext})$)/$R_{min}(B_{ext})$. The authors observed a switching between a finite-resistance and a null-resistance state (i.e., an infinite MR based on their MR definition) when their measurement electrodes were placed at the CrSBr/NbSe$_2$ edge. This behaviour was attributed to a modulation of superconductivity induced by the stray fields generated by CrSBr. The authors also made a lateral superconducting pseudo-spin valve device with two separated CrSBr vdW flakes of different thicknesses, both placed onto a NbSe$_2$ vdW flake. Unlike for a spin valve, where the exchange bias of an additional antiferromagnetic layer is used to pin one of the two Fs to make it harder to switch in an applied $B_{ext}$ (compared to the other F layer), in a pseudo-spin valve the two Fs are chosen to have intrinsically different coercive fields (i.e., different switching fields) [75]. Therefore, in ref. [74] the different (thickness-dependent) coercive fields of the CrSBr flakes allowed to switch their magnetisations from a parallel (P) to an antiparallel (AP) alignment with an applied $B_{ext}$. As a result, the device's resistance could be switched across multiple states, each visible as a plateau in the $R(B_{ext})$ curve measured for the device.

In another study [76], a FGT/NbSe$_2$/FGT (F/S/F) vdW pseudo-spin valve was fabricated, with measurement electrodes placed in contact with the two FGT flakes sandwiching the NbSe$_2$ layer (figures 3(A) and B). As shown in figures 3(C) and (D), below a NbSe$_2$ thickness of ~ 25 nm, the authors observed a 17-fold enhancement in MR (measured in an out-of-plane $B_{ext}$), where MR is defined as MR = ($R_{AP}$ - $R_P$)/$R_P$ ($R_{AP}$ and $R_P$ being the resistances measured in the AP and P state of the two magnetizations of the FGT layers, respectively). Based on a systematic investigation of the MR dependence on $T$ and bias current, also for the devices with different NbSe$_2$ thicknesses, the authors concluded that the MR enhancement results from spin splitting in NbSe$_2$ induced by the PE with FGT. When the NbSe$_2$ is thinner and its superconductivity therefore weaker, the spin-up and spin-down DoS in NbSe$_2$ below $T_c$ are split in energy and have different Δ (Δ$_↑$ for spin-up and Δ$_↓$ for spin-down quasiparticles). At $T \sim T_c$, assuming the magnetisation of FGT is aligned upwards, in thin NbSe$_2$ only Δ$_↑$ is present, whilst Δ$_↓$ is suppressed by the PE (figure 3(E)). This effect allows spin-down electrons from the FGT layer (to which electrical contacts are made for the application of the bias current) to tunnel directly into states at the Fermi level of NbSe$_2$. As a result, the spin polarisation of the current increases and this in turn enhances the MR. Below $T_c$, Δ$_↓$ becomes non-zero, meaning that spin-polarised electrons from FGT can only be injected into the corresponding spin bands of NbSe$_2$ via tunnelling into spin-polarised quasiparticle states above the superconducting gap (figure 3(E)). This leads to a small MR below $T_c$, which causes a significant difference compared to the large MR measured above $T_c$ for a device with thin NbSe$_2$.

When the thickness of NbSe$_2$ increases, the spin-up and spin-down DoS in NbSe$_2$ are degenerate with $\Delta_\downarrow = \Delta_\uparrow$ (since the PE is weaker and superconductivity in NbSe$_2$ is more robust). The MR then becomes less $T$-dependent, and its changes are mainly due to the voltage $V$ applied to the FGT layer. At $T \sim T_c$, electrons from FGT cannot be injected but can only tunnel into the (degenerate) quasiparticle states above the gap in NbSe$_2$ leading to a negative MR because the tunnelling probabilities are set by the spin polarisation of electrons in the FGT layer (figure 3(F)). Below $T_c$, only at a sufficiently high $V$, the chemical potential of the FGT shifts above $\Delta$, which allows electrons to co-tunnel into NbSe$_2$ without altering their spins' polarisation thus making MR not change significantly below $T_c$.

Other groups have shown that the manipulation of the magnetisation of a vdW F can be used to vary the superconducting properties of a vdW JJ embedding such F as a weak link (figure 4(G)). Huang and co-workers [77], for example, studied NbSe$_2$/Cr$_2$Ge$_2$Te$_6$/NbSe$_2$ (S/FI/S) vdW JJs, and chose Cr$_2$Ge$_2$Te$_6$ (CGT) as their FI weak link due to its magnetic structure with magnetic domain walls and domains having comparable sizes (~ 29 nm for domain walls and ~ 100 nm for domains; refs. [78,79]). This peculiar magnetic spin texture of CGT enables the modulation of both in-plane and out-of-plane components of its magnetisation under a small applied $B_{ext}$ (of the order of few millitesla). After measuring the resistance across the JJ ($R_{SFS}$) in an out-of-plane $B_{ext}$ ($B_{out}$) and at different current bias ($I_{bias}$), Huang and co-workers observed a complex modulation of $R_{SFS}$, with dips at small $B_{out}$ (< 10 mT) and for $I_{bias}$ > 150 μA. These dips were attributed to an enhancement in the $I_c$ of the JJ achieved for a specific magnetic configuration of CGT [77]. By analysing the voltage versus current ($V(I)$) characteristics of the JJ, the authors defined two distinct $I_c$ values associated with CP tunnelling through CGT domains with in-plane or out-of-plane magnetisation, respectively (figure 4(H)). The two $I_c$ values exhibit a different dependence on $B_{ext}$ and can be enhanced for specific magnetic configurations (figures 4(J) and (K)), consistent with the observed dips in $R_{SFS}$.

The results in ref. [77] also align with earlier work conducted by Idzuchi and co-workers [78] who studied SQUIDs embedding vdW JJs of NbSe$_2$ flakes coupled via a monolayer of CGT. In these SQUID devices, two distinct $I_c$s were measured each oscillating with a different phase as a function of the flux $\Phi$ through the SQUID loop, although with phase neither equal to 0 nor to π. For conventional S/F/S JJs, an arbitrary phase φ (between 0 and π) can be realised by combining laterally-connected 0- and π-junctions, if the F barrier thickness along the supercurrent flow exceeds the Josephson length $\lambda_J$ [80,81]. $\lambda_J$ defines the length scale over which the supercurrent decays inside the weak link of a JJ [59,82,83]. However, in ref. [78], the CGT thickness is just equal to one unit cell, whilst the estimated $\lambda_J$ is ~ 10 μm. According to the authors, this suggests that the arbitrary φ observed in their SQUID device requires an alternative mechanism as explanation. Idzuchi and co-workers in fact proposed that the phase shift between the NbSe$_2$ layer is determined by the magnetisation direction of the CGT domains [78]. For magnetic domains in CGT with out-of-plane magnetisation, Ising CPs in NbSe$_2$ can tunnel across CGT without spin flipping, thus forming π-phase junctions if the magnetic scattering is strong enough to invert the sign of the tunnelling pairs' wavefunction. In contrast, for magnetic domains with in-plane magnetisation, spin flipping can occur during tunnelling leading to a ground state with a 0-phase shift.

The argument made to explain the above observations in ref. [78] also agrees with that reported in ref. [84], where a SQUID device embedding both a $NbSe_2/CGT/NbSe_2$ JJ and another JJ with two $NbSe_2$ flakes was characterised. While the $NbSe_2/NbSe_2$ JJ shows $I_c$ oscillations in an out-of-plane $B_{ext}$ with a maximum at $B_{ext} = 0$, the JJ with a CGT weak link exhibits a phase shift $\varphi \sim 148.6°$ at $B_{ext} = 0$. Additionally, the temperature dependence of $I_c$, $I_c(T)$, for the JJ with CGT deviates from that described by the Ambegaokar–Baratoff theory [85], according to which $I_c(T)R_N = (\pi\Delta(T)/2e)\cdot\tanh[\Delta(T)/2k_BT]$, where $R_N$ is the normal-state resistance of the JJ and $e$ is the electron charge. Instead, $I_c(T)$ shows a non-monotonous trend, which is attributed to an incomplete 0-to-$\pi$ phase transition leading to an arbitrary phase shift in the CPR [84].

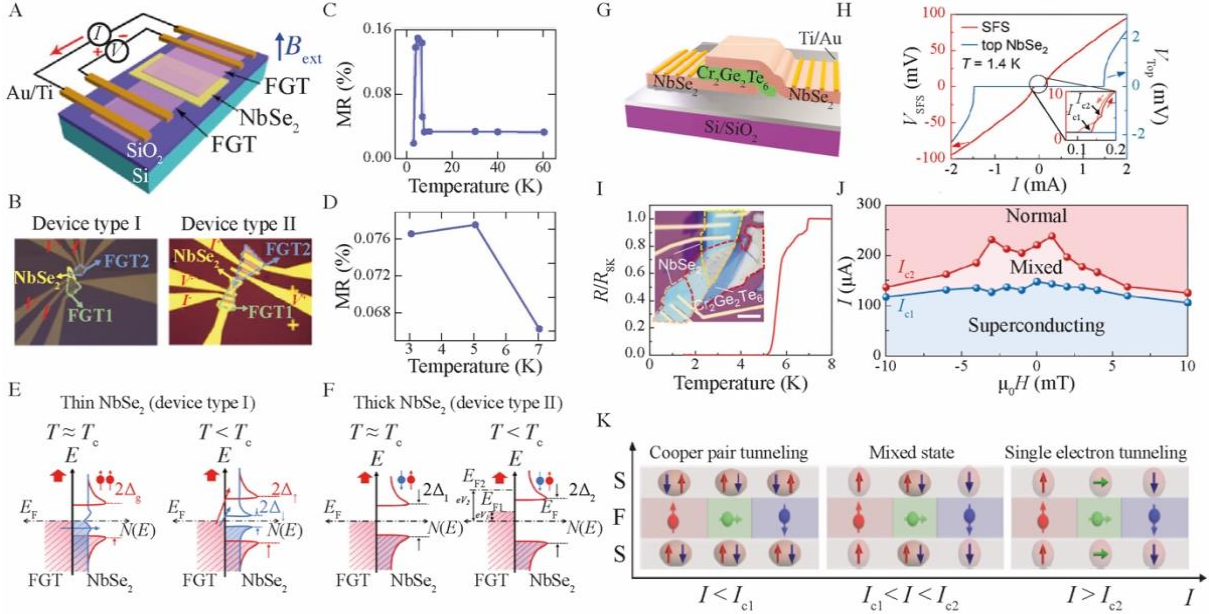

**Figure 3.** From spin-valve behaviour to magnetic domain-engineered supercurrents in vdW devices. (A)-(F) $Fe_3GeTe_2/NbSe_2/Fe_3GeTe_2$ vdW spin valve: (A) device schematic; (B) optical images of two devices with thin $NbSe_2$ (type I, left) and thick $NbSe_2$ (type II, right); magnetoresistance MR versus temperature $T$ for a type-I (C) and type-II device (D) and corresponding band diagrams in (E) and (F), respectively, both for $T$ close to the superconducting transition temperature $T_c$ and below it, illustrating the operation of the devices in each regime (FGT magnetisation assumed oriented upward). (G)-(K) $NbSe_2/Cr_2Ge_2Te_6/NbSe_2$ vdW JJ ($Cr_2Ge_2Te_6$ = CGT): (G) schematic and (I) optical image of the device with resistance versus temperature, $R(T)$, normalised to $R$ in the normal state at 8 K (inset); (H) two-step voltage versus current, $V(I)$, curve measured across the JJ at 1.4 K with critical currents $I_{c1}$ and $I_{c2}$ (red) and $V(I)$ measured for a single $NbSe_2$ flake at the same $T$ (blue); (J) field dependence of $I_{c1}$ and $I_{c2}$; (K) different tunnelling regimes with CPs tunnelling through both domain walls and magnetic domains for $I < I_{c1}$ (left panel), pairs tunnelling only through domain walls and single electrons tunnelling through domains for $I_{c1} < I < I_{c2}$ (middle panel), only single electron tunnelling occurring for $I > I_{c2}$ (right panel). Panels (A)-(F) adapted from [76]; panels (G)-(K) from [77].

In addition to $T_c$ or $I_c$ modulations in a vdW S induced by changes in the magnetization of the vdW F coupled to it, other effects already known for 3D S/F devices have also been reproduced for S/F vdW devices. One of these effects is the transition from a 0- to a $\pi$-coupling between two vdW S layers in a S/F/S JJ, which occurs as the thickness of the vdW F weak link ($d_F$) is progressively varied. The transition provides evidence for the formation of Fulde-Ferrel-Larkin-Ovchinnikov (FFLO) states [59,86,87] at the S/F interfaces of the JJ, which induce an oscillation in the sign of the superconducting order parameter as a function of the distance from the S/F interface (figure 4(A)). In ref. [88], the authors fabricated and tested several $NbSe_2/CGT/NbSe_2$ (S/F/S) vdW JJs, where they systematically changed $d_F$ (figures 4(B) and (C)). They found that the $I_c$ of the JJ drops at nearly zero for a CGT thickness $d_F$ of ~ 8.4 nm, then increases as $d_F$ is increased, before vanishing again at $d_F$ ~ 12.3 nm (figure 4(D)). At a specific $d_F$ of ~ 9.9 nm, the authors also observed an unusual $I_c(B)$ pattern, differing from the Fraunhofer shape due to both a shift of its main maximum from $B_{ext} = 0$ and to its dependence on the $B_{ext}$ history. The latter observation was interpreted by Kang and co-workers as a signature for the coexistence of spatially-separated regions with 0- and π-phase differences, consistent with the findings in ref. [78].

Recent theoretical studies also made interesting predictions on PEs that can only be observed in S/F vdW systems, meaning that they do not have a counterpart in 3D S/F systems. These "vdW S/F PEs" are due to the high responsivity of vdW materials to electrostatic gating. Bobkov et al. [89], for example, suggested that, by tuning the interface hybridisation in a S/F vdW stack via electrical gating, it may be possible to modulate the strength of the PE. If verified, this could enable the realisation of S/F superspintronic or caloritronic devices with functionalities not seen in 3D S/F thin film heterostructures – in 3D S/F systems the strength of the PE is in fact univocally determined by the physical properties of the S and F materials and by their interface transparency [59].

Another unique feature of a PE emerging at a S/F vdW interface has been highlighted in ref. [90]. In a S/F vdW system where F is a vdW bilayer, CPs can be not only homogeneously distributed over both S and F materials – this is the situation described by the mesoscopic FFLO states forming at 3D S/F interfaces – but can be also localised within the same single layer of the vdW F (forming a "local CP") or split between its two layers (forming a "non-local CP"), as shown in figures 4(F) and (G). These non-local CPs are sensitive to the chemical potential of the monolayer on which each electron is localised. The FFLO state formed by such non-local pairs could be tuned by applying a $V_G$ to one of the two monolayers of the vdW F. In a S/F/S vdW JJ with a gated F vdW weak link, the $V_G$-tuneable FFLO state would allow switching of the JJ between 0- to π-coupling purely electrically (i.e., without the need for an applied $B_{ext}$). The realisation of such state could have a profound impact on development of highly-scalable superconducting vdW logic devices and (volatile) vdW memories based on a reversible $V_G$-driven switching between 0- and π-coupling.

The generation of long-ranged spin-triplet pairs at S/F vdW interfaces has also been recently claimed by different groups who have performed studies on both S/F vdW stacks and on S/F/S vdW JJs. The evidence reported in these studies suggest that the generation of spin-triplet CPs does not necessarily require an S/F interface featuring strong (ionic or covalent) bonds like those of a 3D S/F thin film heterostructure [91-94], but can also occur across weakly-bonded vdW interfaces [10-11].

In ref. [10], the authors fabricated lateral $NbSe_2$/FGT/$NbSe_2$ vdW JJs and measured Josephson coupling up to a separation of the $NbSe_2$ flakes across the FGT weak link of ~ 300 nm (figures 4(H)-(J)), which is well-beyond the coherence length for spin singlets in FGT (~ 3.5 nm) [10]. However, the source of magnetic inhomogeneity – typically required in 3D S/F/S JJs to generate long-ranged spin-triplet supercurrents – in the FGT layer remains unclear. The authors argue that a non-coplanar spin texture of FGT, wherein Fe atoms are arranged to form a frustrated triangular lattice, can be at the origin of spin-triplet generation – such arrangement of Fe atoms was already proposed from magnetotransport measurements on FGT in the normal state [95].

Although other studies have also shown that non-coplanar spin textures can generate long-ranged supercurrents – which are possibly carried by spin-triplet CPs – even in antiferromagnets [96], further experiments are needed to verify whether a frustrated Fe lattice in FGT can produce a spin-triplet supercurrent. An alternative explanation made by the same authors in ref. [10] is the formation of 2D superconductivity at the NbSe$_2$/FGT interface, which had been proposed earlier also by Fu and Kane [97]. However, this hypothesis also remains speculative at present and requires direct spectroscopic verification.

Another study on S/F vdW stacks showed indirect evidence for long-ranged spin-triplet generation based on magnetotransport measurements [11]. Here, the researchers used NbS$_2$ as vdW S, and a non-collinearly ordered magnetic material, Cr$_{1/3}$NbS$_2$, as F [98] (figure 4(K)). Cr$_{1/3}$NbS$_2$ belongs to a class of magnetic materials obtained from vdW Ss (e.g., TaS$_2$, NbS$_2$, NbSe$_2$ etc.) via intercalation of transition metals (usually Mn or Cr) [99,100]. Although these materials are not vdW because the transition metal atom increases the strengths of the chemical bonds between consecutive layers in the stack of the original S TMD, they are metallic and can exhibit a non-collinear (helimagnetic) spin texture. As a result, they constitute an interesting system to test spin-triplet generation at S/F vdW interfaces.

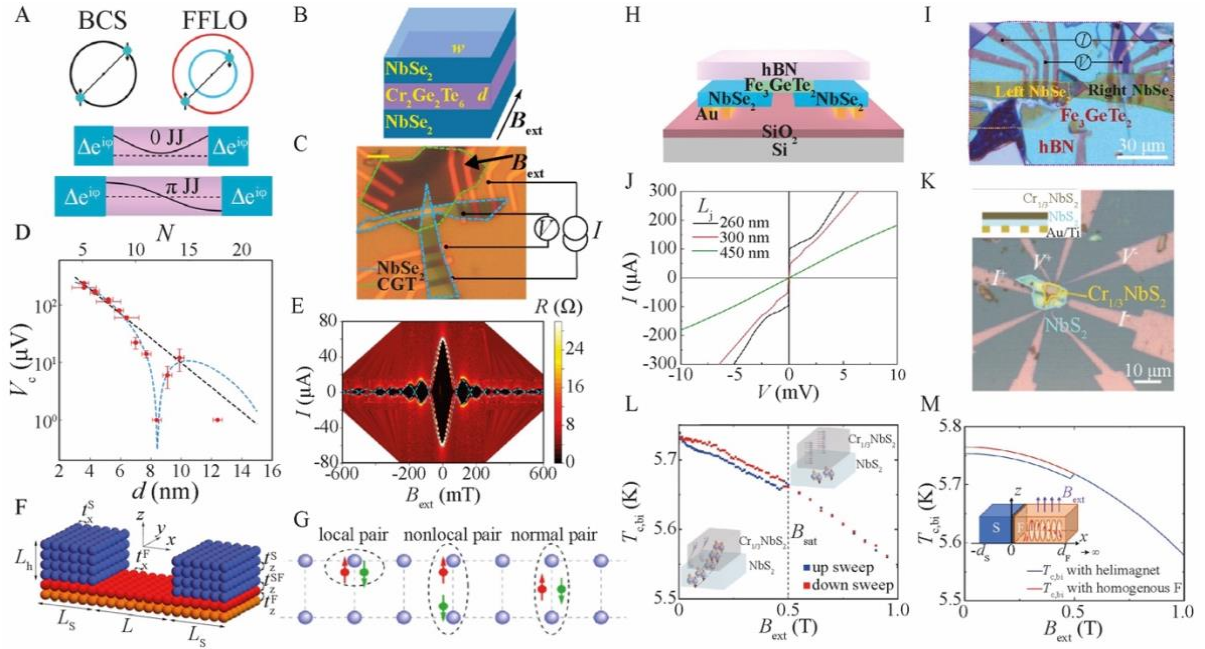

**Figure 4.** Superconducting proximity effects in vdW devices. (A) Pairing schematics conventional spin-singlet pairs according to BCS theory (left) and pairs associated with FFLO states (right) resulting in 0- and π-coupling in S/F/S JJs as the F thickness varies (bottom panel). (B)-(E) S/F/S vdW JJs with NbSe$_2$ electrodes and CGT barrier of thickness $d$ and width $w$: (B) layout and orientation of in-plane field $B_{ext}$; (C) optical image of a JJ device and (D) characteristic voltage $V_c$ versus $d$; (E) differential resistance versus bias current $I$ and $B_{ext}$ for a JJ $d$ = 3.6 nm. (F)-(G) Model of a S/F/S vdW JJ with a bilayer vdW F as weak link and hopping terms $t$ between different layers (F) and classification of proximity-induced pairs in the vdW F as 'local' and 'non-local' pairs (G). For large interlayer hopping in F ($t_z^F$) the pair wavefunction is homogeneous reproducing the conventional proximity regime with an isotropic F. (H)-(J) Lateral NbSe$_2$/FGT/NbSe$_2$ (S/F/S) vdW JJ: (H) schematic; (I) optical image of device; (J) $I$-$V$ curves measured for devices with different separation between electrodes $L_J$ as specified in the legend. (K)-(M) Cr$_{1/3}$NbSe$_2$/NbS$_2$ (F/S) vdW stack: (K) optical image of device; (L) superconducting critical temperature, $T_{c,bi}$, versus $B_{ext}$ for increasing (blue curve) and decreasing $B_{ext}$ (red) in; (M) calculated $T_{c,bi}(B_{ext})$ profile for helimagnetic F (blue) versus homogeneously magnetised F (red). Panels (A)-(E) adapted from [88], (F)-(G) from [90], (H-J) from [10] and (K)-(M) from [11].

In their study [11], Spuri and co-workers tracked the $T_c$ of the $Cr_{1/3}NbS_2/NbS_2$ stack as a function of an applied in-plane $B_{ext}$, which progressively turns the $Cr_{1/3}NbS_2$ magnetic spin texture from helical (inhomogeneous) to ferromagnetic (homogeneous). They observed a positive $T_c$ shift when transitioning from a helical to a ferromagnetic state of the $Cr_{1/3}NbS_2$ flake, which is consistent with spin-triplet generation (figures 4(L) and (M)). This is because, in the helical state, triplet CPs penetrate deeply into the $Cr_{1/3}NbS_2$ layer, thus draining the superconducting $NbS_2$ out of pairs and reducing its $T_c$. When $Cr_{1/3}NbS_2$ is made magnetically homogeneous (i.e. ferromagnetic) by the applied $B_{ext}$, triplet generation is suppressed, which confines CPs at the S/F vdW interface and increases $T_c$. The study of Spuri et al. highlights the potential of helimagnetic metals like $Cr_{1/3}NbS_2$ coupled to vdW Ss for the realisation of bistable superconducting devices where triplet generation can be reversibly switched on/off by an applied $B_{ext}$.

## 5. Quantum sensing and microwave technologies with vdW Josephson devices

Josephson coupling has already been observed in a variety of vdW heterostructures embedding different types of vdW weak links. Unlike JJs based on 3D thin-film stacks, in vdW systems Josephson coupling can be also achieved by stacking two S flakes and using the vdW gap between them as an insulating barrier. This was demonstrated, for example, in $NbS_2/NbS_2$ vdW JJs [101], where Josephson coupling and a Fraunhofer pattern were measured, as well as in $Bi_2Sr_2CaCu_2O_{8+\delta}/Bi_2Sr_2CaCu_2O_{8+\delta}$ vdW JJs [102] where two degenerate Josephson ground states related by time-reversal symmetry could be observed at a twist angle between the flakes of ~ 45°.

Similar results were also reported for $NbSe_2/NbSe_2$ JJs in ref. [15]. Here, the evolution of the $V(I)$ characteristics as a function of the twisting angle θ formed by the in-plane crystallographic axes of the two $NbSe_2$ flakes was studied. A hysteresis in the $I(V)$ curve was seen at θ = 0°, which disappeared as θ increased between 20° and 40°, and reappeared at θ = 60°. This dependence was attributed to the anisotropy of the superconducting order parameter of $NbSe_2$ showing maxima at 60° intervals in its Fermi surface, as evidenced by angle-resolved photoemission spectroscopy (ARPES) measurements [103,104]. In ref. [15], the authors also fabricated a JJ with an optimal misalignment angle θ to get an overdamped junction (i.e., a JJ with no hysteresis in its $V(I)$) and used this as a starting point to realise a SQUID with good performance. Nonhysteretic junctions are advantageous for the fabrication of SQUIDs that can operate with flux-locked loop feedback to reduce their flux noise. A large voltage modulation is essential to minimise the noise from the amplifier in the SQUID feedback loop since the contribution of the amplifier to the noise, $S_{V,a}^{\frac{1}{2}}$, directly affects the flux noise of the SQUID $S_{\Phi,a}^{\frac{1}{2}}$ according to the relation $S_{\Phi,a}^{\frac{1}{2}} \approx \Phi_0 S_{V,a}^{\frac{1}{2}}/\pi\Delta V$ [105]. The vdW-based SQUIDs in ref. [15] showed relatively large voltage modulations up to 1.4 mV under a bias current of 50 μA.

In a very recent study, González-Sánchez and co-workers [16] studied vdW JJs having $NbSe_2$ as vdW Ss and the vdW AFI $NiPS_3$ as weak link (figures 5(A) and (B)). In these devices, the interplay between superconductivity and the magnetic spin texture of $NiPS_3$ leads to the emergence of localised (edge) states that dominate the electronic transport. The interference between the localised states makes the JJ behave like a SQUID, as evidenced by the $B_{ext}$-induced modulation of $I_c$ which resembles that typical of a direct current (DC) SQUID. The SQUID behaviour persists up to in-plane $B_{ext}$ as large as 6 Tesla as shown in figure 5(C), which is well above the maximum $B_{ext}$ reported in the literature for SQUIDs based on 3D Ss [106]. The results obtained by González-Sánchez et al. [16] thus demonstrate a novel simple process for the realisation of SQUIDs that is made possible by the properties of the vdW materials used. The standard approach for creating a SQUID with 3D materials in fact consists in patterning a micron-scale superconducting loop that incorporates at least one JJ and it involves several fabrication steps. In contrast, the approach demonstrated in ref. [16] requires no lithography steps to define the SQUID geometry, but it only involves the stacking of vdW materials. As a result, SQUID devices with significantly smaller footprint and easier to scale up compared to existing SQUIDs can be obtained with the right type of vdW heterostructures. The increased scalability and simpler fabrication of vdW-based SQUIDs may have profound implications across the wide range of fields where SQUIDs are used, which span from superconducting electronics to quantum sensing to medical imaging [107].

VdW JJs have also been studied for the realisation of Josephson parametric amplifiers (JPA) [108]. JPAs are used, amongst other things, in superconducting quantum circuits to amplify the low-power microwave signals used to measure the qubit state. JPAs are typically made using Al/AlO$_x$/Al tunnel JJs in a SQUID geometry, where the enclosed magnetic flux Φ is used to tune the JPA operating frequency [109]. Using vdW heterostructures, it has been shown that it is possible to realise JPAs where the operational frequency can be tuned via a $V_G$ applied to the weak link of one of the SQUID's JJs. In ref. [19], the authors demonstrated a JPA based on a MoRe/graphene/MoRe JJ, where the graphene layer had a thickness of few layers. The JPA was realised in the form of a Duffing LC oscillator [110] consisting of a coplanar waveguide of MoRe with two Al/Al$_2$O$_3$/MoRe capacitors in series that were terminated at the MoRe/graphene/MoRe JJ acting as the inductor of an LC oscillator (figures 5(D) and (E)). The study [19] showed that the performance of the JPA (with a tuneable frequency obtained through $V_G$ application to graphene) is limited only by quantum noise. At the optimal pump frequency and power, the JPA showed a 24 dB gain and a -10 MHz bandwidth (figure 5(F)), which is comparable to the best JPAs previously reported in the literature [108].

VdW materials have also been successfully used for bolometers, which are devices that sense incoming radiation and particles. A common readout scheme in a bolometer consists in monitoring a change in resistance due the thermal energy adsorbed by the device [111]. Graphene, for example, is an ideal vdW material for bolometers due to its weak electron-to-photon coupling and small heat capacity. These properties ensure that electrons in graphene remain thermally isolated from its lattice, which allows energy absorbed from incoming photons or particles to quickly re-equilibrate. As a result, graphene-based bolometers offer very high sensitivity and fast response times. In ref. [17], the authors reported a bolometer made of a NbN/graphene/NbN JJ embedded in a microwave resonator. This bolometer shows a noise-equivalent power of 7 x 10$^{-4}$ pW/$\sqrt{Hz}$ corresponding to the resolution for a single 32-gigahertz photon – which represents the ultimate resolution limit set by intrinsic thermal fluctuations at the device's operating temperature of 0.19 K.

In a different study, Shein et al. [18] also tested superconducting hot-electron bolometers based on NbSe$_2$ (figures 5(G)-(I)). Similar to traditional bolometers made from NbN thin films and operating in the THz range, the NbSe$_2$ bolometers can detect THz radiation over a wide frequency range (from 0.13 to 2.5 THz). By optimising the interface transparency between NbSe$_2$ and the metal leads, the authors could achieve a noise-equivalent power of 7 pW/$\sqrt{\text{Hz}}$ and a nanosecond response time [18].

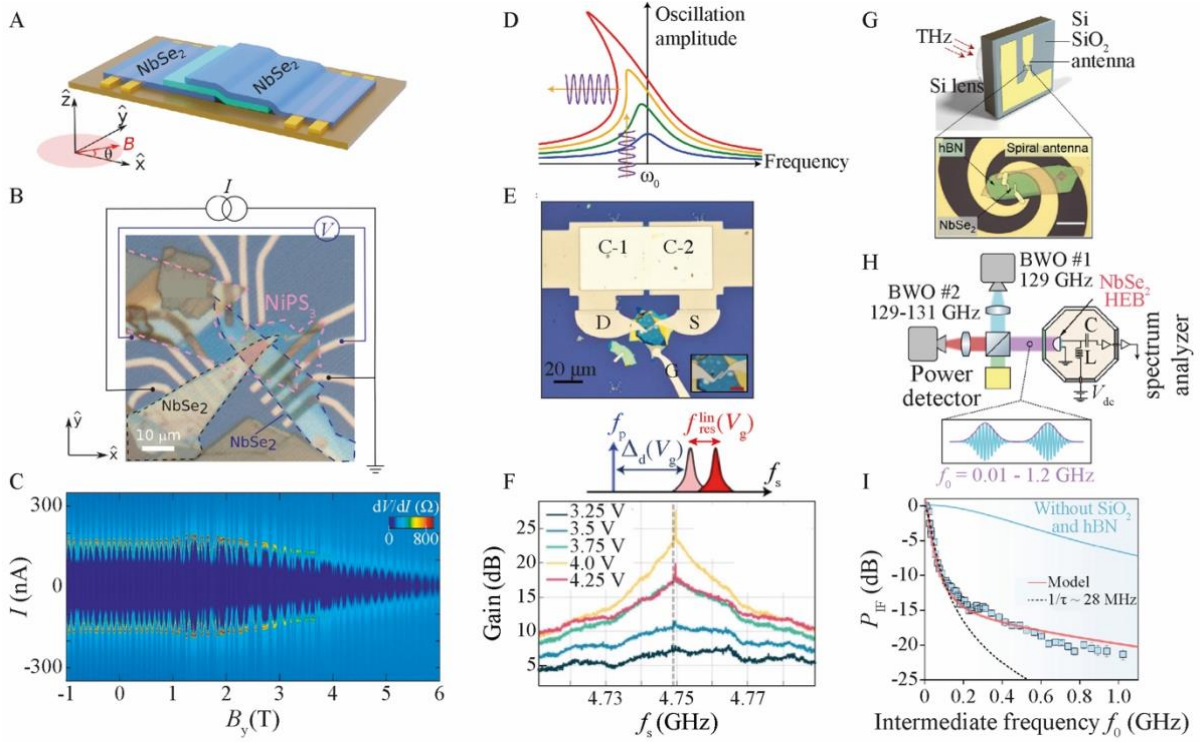

**Figure 5.** Functionalities of devices based on vdW Josephson junctions and superconductors. (A)-(C) Vertical JJ with superconducting NbSe$_2$ electrodes and antiferromagnetic NiPS$_3$ weak link: (A) schematic and in-plane field orientation ($B_y$); (B) optical image of the device; (C) differential resistance, d$V$/d$I$, versus bias current $I$ and $B_y$ showing a SQUID-like interference pattern. (D)-(F) Duffing-oscillator parametric amplifier: (D) schematic of amplitude response versus drive frequency ($x$-axis) and power (curves with different colours); (E) optical image of MoRe/graphene/MoRe JJ on hBN/graphene/hBN (hBN = hexagonal boron nitride) wire bonded to series MoRe-Al$_2$O$_3$-Al capacitors (C-1 and C-2) forming an LC resonator; (F) gain versus probe frequency $f_s$ for different applied gate voltages $V_g$. A fixed pump $f_p$ is used whilst $\Delta_d = f_{\text{res}}^{\text{lin}} - f_p$ is tuned by $V_g$ ($f_{\text{res}}^{\text{lin}}$ is the linear resonance frequency of the device). (G)-(I) Antenna-coupled NbSe$_2$ bolometer: (G) schematic and optical image of the device (50-nm-thick hBN capping; Al contacts; scale bar of 20 μm); (H) heterodyne setup with two backward wave oscillators (BWOs) of which BWO1 is used as local oscillator and BWO2 is used as source of the radiofrequency signal in the 129-131 GHz range, with device mounted on a hemispherical Si lens in a cryostat; (I) output power versus intermediate frequency $f_0 = |f_{\text{BWO1}} - f_{\text{BWO2}}|$ with curves for a simple relaxation model with zero fitting parameters (pink), for Lorentzian decay with 3dB roll-off and inverse relaxation of 28 MHz (black dashed) and for predicted response of NbSe$_2$ directly coupled to the bath (blue). Panels (A)-(C) adapted from [16], (D)-(F) from [19], and (G)-(I) from [18].

## 6. Superconducting devices based on molecules/vdW hybrids

Molecules adsorbed on vdW Ss offer interesting opportunities to widen the control modalities and functionalities of vdW-based superconducting devices. As reported in a recent review article on hybrid molecules/vdW material systems [112], unlike inorganic vdW systems where the physical properties are determined by the atomic arrangements in the crystal lattice, the almost unlimited degrees of freedom in molecular design offer the possibility to synthesize molecules with optically, magnetically or electrically active functional groups that can change the properties of the vdW material to which the molecules are bound.

Studies on molecular layers adsorbed on the surface of an ultrathin S thin film had already shown that molecules can significantly affect the superconducting properties. For example, Yoshizawa et al. [113] investigated heterostructures composed of two types of phthalocyanine molecules – each coordinating a different metal ion (Mn and Cu) – which were deposited onto a superconducting In layer. In this study, the authors showed that the $T_c$ of the In layer could be increased or decreased because of the interaction of In with the Cu- and Mn-phthalocyanine, respectively. The suppression of $T_c$ for the Mn case was explained due to the magnetic moments of the molecular $d$-orbitals. This effect was not observed in the Cu-based phthalocyanine, where $d$-orbitals are confined within the molecules, for which a $T_c$ increase was measured due to the $p$-type doping induced by the molecules into the In layer.

Several groups have already started to investigate the effect of molecular deposition on vdW Ss. Zhu et al. [114], for example, showed that the decoration of $NbSe_2$ with hydrazine molecules can induce ferromagnetism on the $NbSe_2$ surface. According to the authors [114], hydrazine triggers an elongation of the Nb-Se bond, which enhances the ionicity of the tetravalent Nb with unpaired electrons, and this in turn results in ferromagnetic ordering.

Calavalle et al. [115] studied the effect of the adsorption of self-assembled monolayers (SAMs) of fluorinated or amine-containing molecules on the superconducting properties of a $NbSe_2$ monolayer. The adsorption of fluorinated-containing molecules of trichloro(1H,1H,2H,2H-perfluorooctyl)silane (TPS) led to an increase in the $T_c$ of $NbSe_2$ by more than 50%, whilst the amine-containing molecules of N-[3-(trimethoxysilyl)propyl]ethylenediamine (AHAPS) led to a decrease in $T_c$ of $NbSe_2$ by more than 70%. By measuring how the work function φ of $NbSe_2$ is modified upon absorption of each type of molecules, the authors found that TPS increases φ by +0.63 eV due to hole doping, whilst AHAPS decreases φ by -0.60 eV due to electron doping. This difference is attributed to the opposite orientation of the out-of-plane permanent electric dipoles of the two molecular layers. The experiment also suggests that, if similar molecules with a permanent dipole were adsorbed onto a $NbSe_2$ monolayer, offering also the possibility to re-orient the dipole with an external excitation (e.g., optical stimuli), one could realise bistable superconducting devices with well-separated $T_c$ values associated to each state.

It was also recently proposed that the incorporation of chiral molecules (ChMs) into layered vdW Ss can create chiral superconductivity with time reversal symmetry breaking (TRSB) [116]. Ss with a chiral order parameter and TRBS intrinsically have CPs of electrons in a spin-triplet state [117], which is of course interesting not only from a fundamental perspective but also because they can be used as an intrinsic source of spin-triplet supercurrents (i.e., not requiring coupling to a F layer) in superspintronic devices. Chiral superconductors with TRSB, however, are rare and include only few examples of Ss like $Sr_2RuO_4$ and U-containing compounds like $UPt_3$ and $UTe_2$ [118-122]. For $Sr_2RuO_4$, recent experiments have also questioned the TRSB nature of its superconducting order parameter [123].

A recent study on ChMs-intercalated TaS$_2$ flakes reported evidence for unconventional superconducting properties [116]. In particular, the authors measured an in-plane $B_{c2//}(0)$ exceeding $B_P(0)$, a robust π-phase shift in the Little-Parks oscillations [124], and an $B_{ext}$-free superconducting diode effect (see section 7), which they interpreted as evidence for spontaneous TRSB.

The Little-Parks effect, namely the oscillation of $T_c$ of a superconducting ring as a function of externally applied magnetic flux $\Phi_{ext}$, originates from the contribution of the screening current to the total fluxoid $\Phi$ and from the quantisation of the latter [124]. This effect translates into the oscillation of the ring resistance close to $T_c$ as function of $\Phi_{ext}$. Fluxoid quantisation implies that the difference between $\Phi$ and $\Phi_{ext}$ is compensated by a spontaneous supercurrent $\vec{j}$. As $\vec{j}$ flows, it induces a change in the system's free energy $F_N \propto \vec{j}^2 \propto (\Phi_{ext} - \Phi)^2$, which is the reason behind the oscillations of $F_N$ with $\Phi_{ext}$. Just below $T_c$, $R$ is sensitive to $F_N$, which makes Little-Parks oscillations resolved in $R$ a tool to probe the phase of the superconducting order parameter. For conventional Ss, $R$ shows a minimum when $\Phi_{ext} = 0$, whereas in some unconventional Ss, this minimum can occur at a non-null $B_{ext}$ value. In ref. [116], the authors observed Little-Parks oscillations in the $R$ of the ring made with ChMs-intercalated vdW TaSe$_2$ with a maximum (other than a minimum) at $B_{ext} = 0$, which hints toan unconventional nature of the superconducting order parameter of the ChMs/TaSe$_2$ hybrid. When the same ring was fabricated using TaS$_2$ without ChMs or using TaS$_2$ intercalated with achiral molecules instead, a minimum in $R$ at $B_{ext} = 0$ (other than a maximum) was observed, thus confirming that an unconventional superconducting state is just observed for the ChMs-intercalated TaS$_2$ system.

## 7. VdW superconducting diodes

Several groups have explored superconducting vdW heterostructures to realise a superconducting diode effect (SDE) [125]. Such a device can be operated as a switch. For instance, if the $I_c$ at positive polarity of $I_{bias}$ ($I_c^+$) is higher than the $I_c$ at negative polarity ($I_c^-$), it is possible to choose a specific $I_{bias}$ value, $\tilde{I}_{bias}$, such that $|I_c^-| < \tilde{I}_{bias} < I_c^+$. At this $\tilde{I}_{bias}$, the state of the device can be reversibly switched from superconducting to resistive by alternating the $\tilde{I}_{bias}$ polarity.

The observation of a SDE usually requires an applied $B_{ext}$, since both inversion symmetry and time-reversal symmetries have to be broken for the SDE to occur. A change in the sign of $B_{ext}$ can be also exploited to change the diode's polarity, which can be defined as the sign of the difference between $I_c^+$ and $|I_c^-|$. To date, the SDE has been observed in several vdW systems including bare vdW S flakes [126], magic-angle-twisted bilayers [127,128], small-twist-angle trilayer graphene [129], and vdW JJs [20].

In ref. [126], for example, Bauriedl et al. patterned constrictions from NbSe$_2$ flakes to ensure a well-defined current density direction and could observe a SDE independently on the number of layers of NbSe$_2$. Although the SDE should not be observed when the number of layers is even because such a system does not break inversion symmetry, the authors argued that, due to the weak vdW interaction between subsequent layers, NbSe$_2$ effectively behaves as a stack of disconnected monolayers – which can explain the occurrence of the SDE in NbSe$_2$ also for an even number of layers. To observe the SDE, Bauriedl and co-workers had to apply an out-of-plane $B_{ext}$ ($B_{ext,out}$), in contrast with 3D Ss with Rashba SOC [130,131] where the SDE is driven by an in-plane $B_{ext}$, $B_{ext,in}$, applied perpendicular to the supercurrent. When a non-null $B_{ext,in}$ was added on top of $B_{ext,out}$, this made the SDE asymmetric in $B_{ext,out}$, meaning that the SDE was enhanced for one $B_{ext,out}$ polarity and reduced for the opposite [126]. According to the authors, this observation suggests that an in-plane component of the SOC plays a crucial role toward the SDE in superconducting TMDs like $NbSe_2$. Such in-plane Rashba-type component of the SOC can also originate from ripples in the TMD flakes [132] or even from the substrate onto which they are transferred [133].

Ref. [134] investigated the SDE in both $NbSe_2/CrPS_4$ vdW stacks and $CrPS_4/NbSe_2/CrPS_4$ vdW superconducting pseudo-spin valves. For a $NbSe_2/CrPS_4$ vdW device, a SDE with rectification efficiency $\eta = ||I_c^-| - I_c^+|/(|I_c^-| + I_c^+)$ up to 10% was achieved and a magnetochiral anisotropy (MCA) $\gamma$ of $\sim 1.5 \times 10^5$ $T^{-1} A^{-1}$ at $T = 1.5$ K was measured. In a device with non-reciprocal electronic transport, the MCA quantifies the strength of the non-reciprocity according to the relation $R = R_0(1 + \gamma I_{bias} B_{ext})$ [135,136]. Metals typically exhibit small MCA values ranging from $10^{-3}$ to $10^2$ $T^{-1} A^{-1}$ [130,135,136] because the SOC and magnetic energies are much lower than the Fermi energy $E_F$, which is of the order of electronvolts, and therefore act as weak perturbations [137]. However, in non-centrosymmetric TMD Ss like $NbSe_2$, the MCA can be enhanced by orders of magnitudes compared to its normal-state value because the role of $E_F$ is played by $\Delta$ below $T_c$ [137], which explains the large $\gamma$ observed in ref. [134]. In the same study made by Yu et al. [134], the authors also reported $\eta$ up to 40% for the $CrPS_4/NbSe_2/CrPS_4$ vdW devices, for which they observed a SDE near $B_{ext} = 0$ with strong oscillations in $\eta$ measured in the low-$B_{ext}$ regime (up to $\sim 0.2$ T). These oscillations were attributed to fluctuations in the magnetic moments of the vdW antiferromagnet $CrPS_4$.

Although a SDE usually requires the application of a $B_{ext}$ for TRSB, the realisation of a $B_{ext}$-free SDE is particularly appealing for applications because it would enable SDE devices that can be controlled without any $B_{ext}$ and in turn remove any possible interference with other superconducting devices that may derive from the $B_{ext}$ applications. The first realisation of a $B_{ext}$-free SDE to date was achieved with a vdW device in ref. [20] (figures 6(A)-(C)). In this work, Wu and co-workers tested $NbSe_2/Nb_3Br_8/NbSe_2$ vdW JJs and observed a $B_{ext}$-free SDE both for $Nb_3Br_8$ weak links consisting of an odd number of layers – where inversion symmetry is by definition broken – and of an even number of layers. $Nb_3Br_8$ is a moderate-gap insulator with a unit cell consisting of six layers of Nb-Br edge-sharing octahedra [138]. Based on their observations, the authors argued that the breaking of inversion symmetry required for the SDE originates from the asymmetry existing between the two $NbSe_2/Nb_3Br_8$ vdW interfaces (e.g., as result of the fabrication process), and therefore is independent of the parity of the number of layers. The combination of such an interfacial asymmetry with the separation between negative and positive charge centres in $Nb_3Br_8$ – which is considered an obstructed atomic insulator [139] where charges are pinned at the unoccupied inversion centres in between the Br-Nb-Br layers – can induce an out-of-plane polarisation and in turn asymmetric Josephson tunnelling leading to the observed $B_{ext}$-free SDE. This argument also suggests that vdW JJs containing vdW materials with an intrinsic polarisation (e.g., ferroelectric or magnetoelectric) as weak links should be investigated in the future for the realisation of a $B_{ext}$-free SDE.

Pal and colleagues [140] also built a JJ where a type-II vdW Dirac semimetal, NiTe$_2$, was used as weak link between superconducting Nb electrodes (figure 6(D)). They observed a SDE, even though NiTe$_2$ is centrosymmetric, which rules out structural asymmetry as the reason underlying the SDE. The non-reciprocal charge transport appeared only with a small in-plane $B_{ext}$ and was maximized when $B_{ext}$ was perpendicular to bias current, with multiple sign reversals as $B_{ext}$ was swept. This phenomenology, together with field-angle systematics and Fraunhofer-pattern shifts, was captured by a model according to which an in-plane $B_{ext}$ imparts a finite centre-of-mass momentum ($q$) to CPs — originating from spin-helical topological surface states of NiTe$_2$ — thereby phase-shifting the CPR and producing diode behaviour. Experiment and theory reported in the paper link the JDE directly to finite-$q$ pairing rather than to heating or geometric rectification (figures 6(E) and (F)) and demonstrate large tuneable asymmetry (up to ~60%) with modest $B_{ext}$. This study [140] highlights topological vdW semimetals as promising materials for the realisation of field-programmable nonreciprocal elements for superconducting electronics.

The SDE can be also exploited to realise a sensing device that works as an antenna, as shown by Zhang and co-workers [141] using a vdW NbSe$_2$ flake. In this study, the authors measured second-harmonic resistance $R^{2\omega}$ of NbSe$_2$ under $B_{ext}$, $R^{2\omega}(B_{ext})$, to verify that it was antisymmetric about $B_{ext} = 0$ – this is equivalent to proving that the device supports a SDE since $\gamma = \frac{2R^{2\omega}}{R_0 I_{bias} B_{ext}}$ [137]. By measuring the dependence of the peak in $R^{2\omega}(B_{ext})$ on the amplitude of an alternating current (AC) $I_0$ injected through the device, Zhang et al. also found that the peak first increased with $I_0$ and then decreased for $I_0 > 20$ µA. Assuming that the SDE originates from vortex motion, the authors argued that the observed dependence in the $R^{2\omega}(B_{ext})$ on $I_0$ is due to a decrease in the pinning potential of vortices at larger $I_0$. Based on these results, they developed a device consisting of a NbSe$_2$ flake acting as an antenna contacted with electrodes to measure the voltage drop $V_{dc}$ and separated from a resistor used to apply an AC signal was developed (figure 6(G)). The device acted as an AC-to-DC rectifier and the voltage drop $V_{dc}$ measured across the device had the same shape as $R^{2\omega}$ (figures 6(H) and (I)). The device also showed broadband sensing from 5 MHz up to 900 MHz (limited only by the upper frequency of the experimental setup), with a maximum response achieved for a frequency of the input AC signal of ~ 200 MHz [141].

Recently a SDE was reported in ultrathin flakes (< 4 unit cells) of the antiferromagnetic Mott insulator α-RuCl$_3$ in proximity coupling to the vdW S NbSe$_2$ [142]. The SDE was observed for a weak out-of-plane $B_{ext}$ with η reaching a maximum of ~ 14% at $B_{ext} = \pm 3$ mT. Since the SDE could be only measured for an out-of-plane $B_{ext}$, the authors argued that the superconducting state in α-RuCl$_3$ must be of Ising type. In the case of Ising superconductivity, an applied out-of-plane $B_{ext}$ should in fact introduce an additional Zeeman shift in the split bands at the K and K' points, which can lead to a SDE. By contrast, if superconductivity in α-RuCl$_3$ were dominated by Rashba SOC due to the breaking of inversion symmetry at the NbSe$_2$/α-RuCl$_3$ vdW interface, an in-plane $B_{ext}$ should introduce Zeeman splitting in the spin bands and hence an SDE, which was not observed.

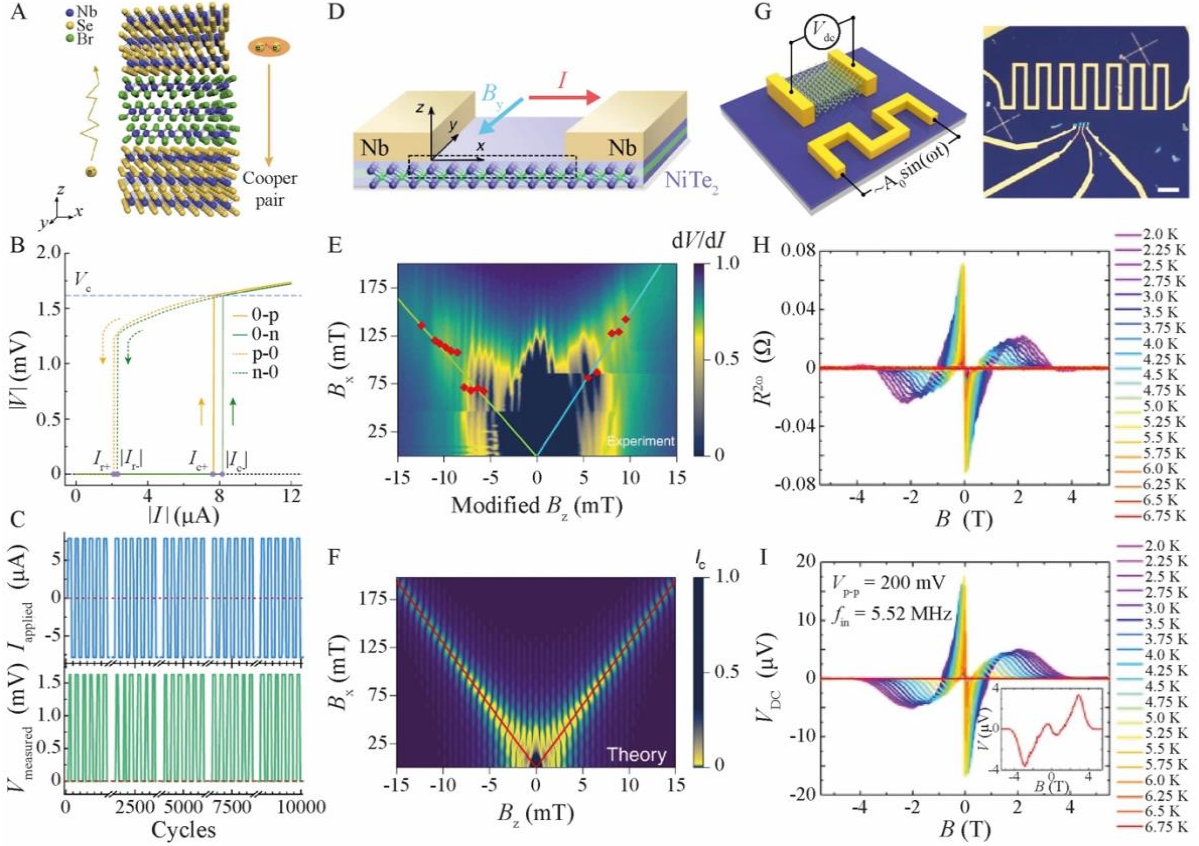

**Figure 6. Schematics and measurements of vdW devices showing a superconducting diode effect (SDE)** (A)-(C) Vertical vdW JJ made of a $NbSe_2/Nb_3Br_8/NbSe_2$ vdW stack: (A) device schematic; (B) amplitude of voltage $|V|$ versus bias current $|I|$ measured for positive (orange) and negative (green) sweeps of $I$ showing a SDE; (C) Square-wave drive $I_{applied}$ alternatively switched between $I_c^+$ and $|I_c^-|$ at 0.5 Hz for 10,000 cycles (upper trace) with the lower trace being the voltage response measured across the JJ. (D)-(F) Lateral JJ with Nb electrodes and a $NiTe_2$ vdW weak link with (E) measured $dV/dI$ versus in-plane $B_x$ and out-of-plane $B_z$ for a 350-nm-wide JJ (geometry in (D)) and (F) calculated interference map (colour given by $I_c$ amplitude). The slope of the line marked in (F) satisfies $\Delta B_x/\Delta B_z \sim 13$, whilst red dots in (E) mark peak centres. The periodicity in $B_z$ in (F) matches that of $I_c$ in (E) and is consistent with CPs acquiring finite momentum in $NiTe_2$ which fixes the SDE polarity. (G) $NbSe_2$ "antenna": an alternating current (AC) is applied to a side resistor whilst a nanovoltmeter measures the direct current (DC) voltage across $NbSe_2$. An optical micrograph of the device is shown at right of panel (G). (H)-(I) Second-harmonic resistance, $R^{2\omega}$, as a function of magnetic field $B$ for the device in (G) with temperature changing from 2 to 6.75 K without any AC drive (H) and with an AC drive (I) having amplitude of 200 mV and frequency of 5.52 MHz. The antisymmetric $R^{2\omega}(B)$ and SDE arise from vortex in an antisymmetric pinning landscape of potentials due to disorder or defects in inversion-broken $NbSe_2$. The similarity in $R^{2\omega}(B)$ and $V_{DC}(B)$ indicates rectification via a net vortex drift under the AC driving force (introduced by the electromagnetic wave radiated through the resistor and sensed by $NbSe_2$). Panels (A)-(C) adapted from [20], (D)-(F) from [140], (G)-(I) from [141].

## 8. Electrostatic tuning of superconductivity and gate-controlled supercurrent devices

The application of a $V_G$ is a well-established approach to trigger superconductivity in semiconducting TMDs. A $V_G$-induced superconducting transition has already been reported for several semiconducting TMDs including $MoS_2$ [143], $WS_2$ [144], $MoTe_2$ [145], and $WTe_2$ [12,13] as well as for twisted bilayer [146] and trilayer graphene [147]. Some research groups have also investigated in more-detail the fundamental properties of the superconducting state induced by $V_G$ in these TMDs. For example, ref. [148] reported measurements of the superfluid response of a $MoS_2$ flake driven superconducting by a $V_G$ (applied via ionic liquid gating) and found signatures consistent with a Berezinskii-Kosterlitz-Thouless (BKT) transition [149,150]. By measuring the superfluid stiffness $\rho_S$ of the MoS$_2$ flake close to its $T_c$ with a scanning SQUID magnetometer, the authors observed a pronounced kink in the magnetic susceptibility, which in turn depends on $\rho_S$, consistent with a BKT transition in the presence of finite-size disorder effects [151]. These findings represent a first step to better understand the nature of the superconductivity induced by $V_G$ in these TMDs. Measurements of $\rho_S$ have in fact already been crucial to establish the unconventional nature of the superconducting state in other 3D Ss [152-154].

A reversible $V_G$ control of superconductivity was also achieved in JJs with a gated graphene weak link – which represent a vdW-based realisation of a class of devices currently known as electrostatically-tuneable Josephson field-effect transistors (JoFETs) [155-157]. In JoFETs, the application of $V_G$ changes the chemical potential of the weak link of a JJ, and in turn affects the superconducting state induced in it. Great advancements have already been made with graphene-based JoFETs. For example, Wang and co-workers [158] showed that graphene-based JoFETs can be integrated into the typical architecture of transmon qubits [159-160] to realise qubits with a $V_G$-tuneable frequency also known as gatemon [161-163] (further details on the application of vdW Ss in transmon qubits are reported in section 10). The spectra of the coplanar waveguide resonators used for the control/readout of these graphene-based JoFETs exhibit features that reflect the electronic transport properties of the graphene layer including voltage tuneability over more than 6 GHz [158], as shown in figures 7(A) and (B).

A different class of $V_G$-controlled devices, known as gate-controlled supercurrent (GCS) devices [164] has also been realised using vdW Ss. Unlike for JoFETs, GCS devices do not rely on the $V_G$-induced modulation of the chemical potential of a semiconducting weak link in a JJ, but rather on an applied $V_G$ to reversibly suppress the $I_c$ of a nanoscale constriction entirely made of a metallic S [164]. In particular, the applied $V_G$ is used to make the device switch from a fully superconducting state (i.e., with $I_c \neq 0$) to a resistive state (i.e., with $I_c = 0$). GCS devices have raised great interest because they are promising for the development of voltage-controlled logics, which would in principle outperform any other existing superconducting logics and be easier to interface with CMOS logics [164].

In ref. [14] a GCS switch based on GaSe/NbSe$_2$ vdW heterostructure was reported. The applied $V_G$ here modulates the ferroelectric polarisation of the GaSe flake, which in turn affects the superconducting properties of NbSe$_2$ including its $T_c$, upper critical field $H_c$ and $I_c$ (figures 7(C)-(F)). The authors ascribe these effects to the leakage current $I_{leak}$ induced by the applied V$_G$, according to the classifications of the possible physical scenarios responsible for the GCS reported in ref. [164]. By properly choosing the working temperature of the device and reversibly switching on/off the applied $V_G$, a switching of the vdW device from a superconducting state to a resistive state could be achieved with a relatively low $V_G$ (~ 4.6 V) [14]. By comparison, the typical $V_G$ values needed to control a GCS device made from a 3D S like Nb are of few tens of volts [165]. However, the device did not remain in the resistive state when the applied $V_G$ was removed, meaning that it did not preserve its state in the remanent polarised state of GaSe [14].

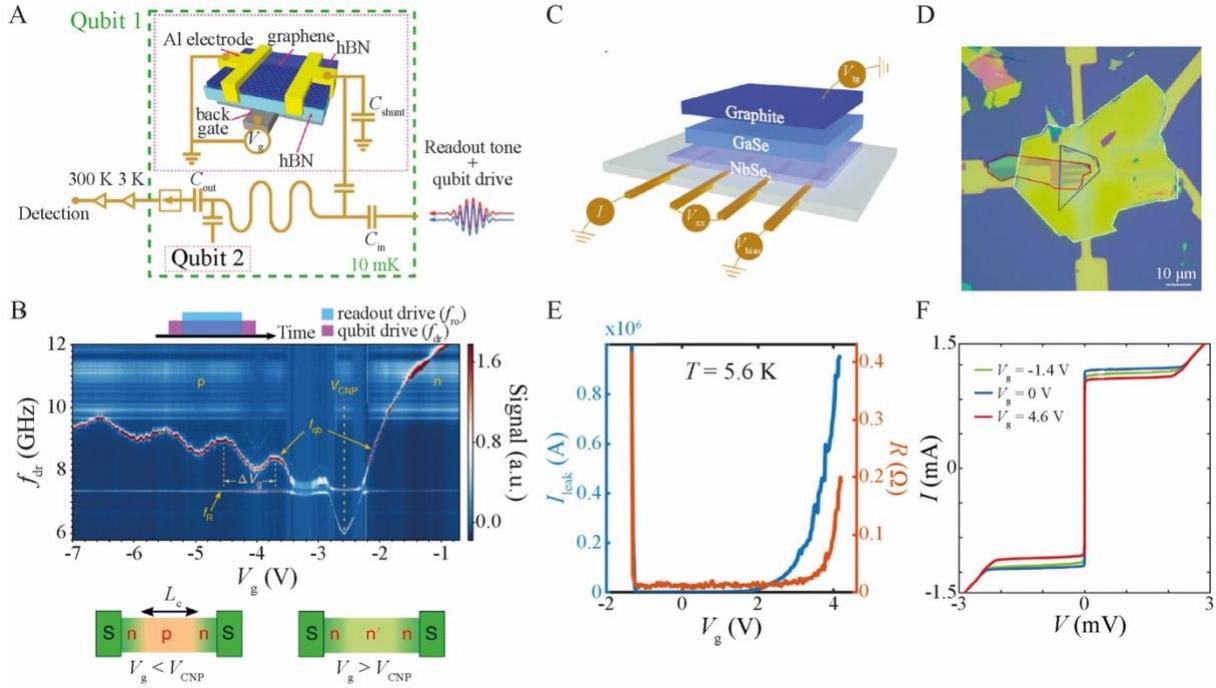

**Figure 7.** Gate-tuneable vdW superconducting devices. (A)-(B) Al/graphene/Al JJ (encapsulated with hBN) embedded in a quantum electrodynamic circuit: (A) device schematic with (B) corresponding qubit spectrum versus frequency drive $f_{dr}$ and the gate voltage $V_g$. The qubit response is asymmetric with respect to the charge neutrality point ($V_g \sim -2.5$ V); in the p-doped regime ($V_g < -2.5$ V), Fabry-Pérot oscillations and aperiodic fluctuations appear. The cavity length $L_c$ extracted from these measurements is $\sim 110$ nm. (C)-(F) Top-gated $NbSe_2$ device: (C) schematic and (D) optical image; (E) Resistance and leakage current $I_{leak}$ versus $V_g$; (F) current versus voltage, $I(V)$, characteristics at representative $V_g$ (values specified in the legend) showing gate-induced suppression of the critical current (f). Panels (A)-(B) are adapted from [158] and (C)-(F) from [14].

## 9. Material engineering aspects and scalability of vdW devices

Most of the studies on vdW Ss have been performed on exfoliated flakes or vdW heterostructures based on them, which are not optimal for technological applications due to their poor stability in air and small lateral size (usually < 100 μm), which limits the number of devices that can be patterned.

The latter issue, meaning the fabrication of large-area vdW Ss, has been addressed in several studies over the past years. Large-area vdW Ss, for example, have been successfully produced by chemical vapour deposition (CVD) [166] and molecular beam epitaxy [167]. In addition, a two-step growth process has been developed which allows to produce stable vdW materials including vdW Ss, halides or carbides over large areas [168]. The idea of this approach is to fabricate the vdW materials through a process that is completely oxygen- ($O_2$) and water- ($H_2O$) free, unlike the CVD typically used. In CVD, oxygen is in fact present in the environment and/or in the precursor gases, which leads to the formation of oxygen bonds to the vdW material. The trapped oxygen can diffuse into the interlayers of the vdW Ss after growth, possibly facilitated by vacancies, which leads to poor air stability of the resulting vdW material. In ref. [168], the authors carried out a two-step process consisting of a physical vapour deposition step (PVD) followed by a CVD step, where $H_2O$ and $O_2$ could be completely removed. To grow air-stable $NbSe_2$, for example, the authors first conducted magnetron sputtering in an ultrahigh vacuum chamber (to deposit the Nb) and then a CVD step to transform Nb into $NbSe_2$ via selenization. The as-obtained $NbSe_2$ films were close to being single-crystalline, although their $T_c$ was lower than that of exfoliated flakes. At the same time, $B_{c2}$ was higher than exfoliated NbSe$_2$ flakes, most likely due to interlayer coupling at the grain boundaries which makes the as-grown NbSe$_2$ not properly vdW. Nonetheless, the NbSe$_2$ thin films obtained showed good stability even after undergoing several treatments such as heating in air at 50 °C for 10 hours or immersion in aqueous solutions including 0.1 M NaOH or 0.1 M HCl solutions, with only minimal changes to their $T_c$ (up to ~ 0.2 K at most).

Another study carried by Zhou et al. reports the wafer-scale growth of vdW heterostructure made of PtTe$_2$/NbSe$_2$/MoS$_2$/WS$_2$ using CVD [169]. During the growth process, the temperature was progressively reduced whilst going from the bottom (WS$_2$) to the top (PtTe$_2$) vdW layer. Transmission electron microscopy and low-temperature transport measurements were done which prove that the vdW heterostructure has sharp interfaces and robust superconducting properties. The approach developed in this study therefore overcomes a problem related to the CVD growth of wafer-scale vdW stacks with vdW Ss, where the lack of stability of the vdW S layer had long represented a major issue.

The non-destructive patterning of vdW Ss into nanoscale devices represents an additional challenge for the fabrication of superconducting devices and nanocircuits based on them. For device applications, it is not only necessary to grow air-stable vdW Ss over large areas, but also that these materials preserve their superconducting properties after patterning. A multi-step patterning process, involving steps like e-beam lithography and reactive ion etching, usually degrades vdW Ss and hence, in turn, their physical properties. Recently, however, a new method based on topotactic fabrication for the patterning of vdW Ss into nanoscale devices was reported [170]. In their study, the authors managed to produce a stable NbSe$_2$ superconducting nanowire single photon detector by first patterning a metallic Nb thin film grown by magnetron sputtering and then by converting the partially oxidised Nb film into NbSe$_2$ via topotactic selenization. For the topotactic selenization, the patterned (partly oxidised) Nb sample and Se powder were placed inside a two-zone furnace, at 800 °C and 450 °C respectively. Here, Ar and H$_2$ were then introduced as carrier gases, and the sample was annealed for ~ 6 minutes. Compared to a top-down approach where the Nb thin film first undergoes selenization to form NbSe$_2$ followed by patterning with electron beam lithography and etching, this fabrication method yielded nanoscale superconducting devices with very good stability in air (figure 8).

## 10. Integration into superconducting quantum circuits

VdW materials have been also studied for the realisation of superinductors or hyperinductors, which are nearly-linear inductive elements used to suppress charge noise in superconducting circuits. Traditionally, superinductors are made using arrays of JJs or nanowires with high kinetic inductance $L_k$ and are key for the fabrication of superconducting qubits protected from noise and with long coherence times [171-173].

$L_k$ depends on the inertia of charge carriers and is inversely proportional to the cross-sectional area of a S, which is one of the main reasons why $L_k$ can be very large in ultrathin vdW Ss like NbSe$_2$ and TaS$_2$. Thanks to their large $L_k$ and low losses in the microwave regime, vdW Ss are quickly emerging as an appealing alternative to the more traditional high-$L_k$ nanowires or arrays of JJs for the realisation of superinductors. Superinductors based on vdW Ss have been already used for the realisation of fluxonium qubits – these are qubits where quantum information is encoded in the magnetic flux passing through a loop of JJs – with reduced loop size [171] and high coherence. Fluxonium qubits can also find applications in quantum circuits for the detection and manipulation of MZMs [174,175].

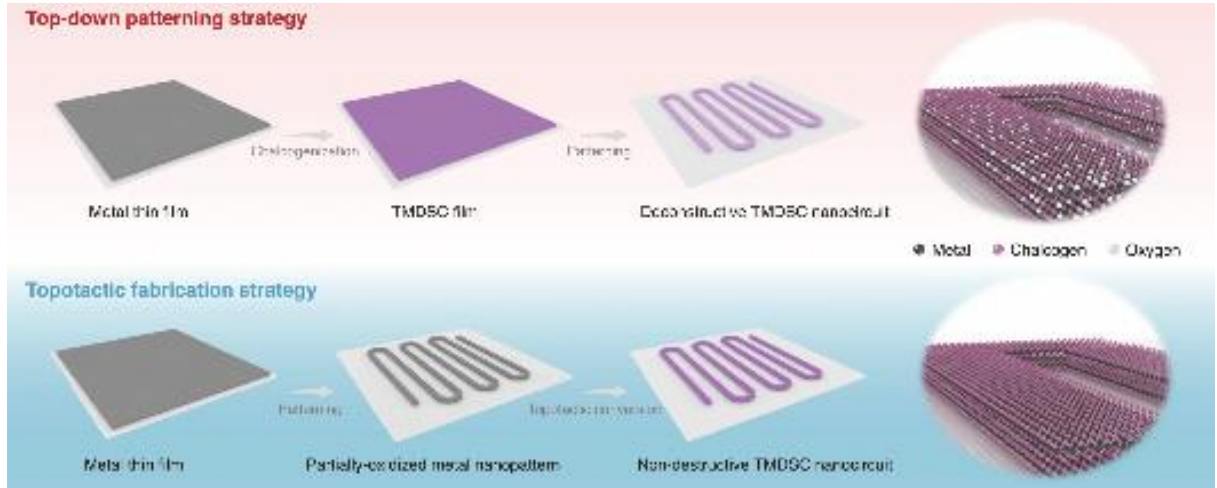

**Figure 8.** Topotactic fabrication of vdW superconducting circuits. The fabrication strategy consists in growing a transition metal dichalcogenide superconductor (TMDSC) by first growing a S thin film, then patterning it into the desired circuit geometry and finally carrying out the topotactic selenization inside a high-temperature furnace (bottom panel). This is different from the usual approach consisting in growing first the S thin film, then carrying out its chalcogenization and finally patterning it into the desired circuit geometry. Figure reproduced from [170].

VdW Ss decoupled by a tunnel barrier made of a hBN or of a semiconducting transition TMD tunnel barrier can be also exploited as an alternative to thin-film multilayers to realise the S/I/S JJs used in another type of superconducting qubits known as transmons – which are charge-based qubits essentially consisting of a JJ with a large shunt capacitor. The use of a vdW flake as the insulating weak link would in fact give a tunnel barrier more stable and flatter than the oxide tunnel barrier (typically $Al_2O_3$) used in conventional transmons. This has already been demonstrated for transmon qubits including S/I/S with Al as S, but where the $Al_2O_3$ (I) layer was replaced by a $MoS_2$ barrier [176]. The as-realised qubits already showed a coherence time of ~ 12 ns, although an additional enhancement of their performance would be doable if $MoS_2$ barriers with fewer defects than those grown by metal-organic CVD (MOCVD) were used. In a more recent study, a transmon qubit embedding a shunt capacitor made of a vdW stack of $NbSe_2$/hBN/$NbSe_2$ was tested [177]. Since the authors used vdW flakes made by mechanical exfoliation, and therefore with fewer defects than MOCVD-grown flakes, they managed to achieve a long coherence time of ~ 1.06 μs. As outlined by the authors of the same study, high-quality vdW heterostructures can help reduce the area of modern qubits by a factor larger than $10^3$, whilst preserving their quantum coherence properties [177].

Recent studies have also suggested qubit architectures based on JJs of twisted vdW flakes of the high-temperature oxide S $Bi_2Sr_2CaCu_2O_{8+\delta}$. For example, in ref. [178] the authors proposed the so-called flowermon qubit design consisting of a vdW cuprate junction with a twist angle close to 45° like those studied in ref. [102] combined with a shunting capacitor. This study suggests that, by properly choosing the twist angle in this type of cuprate vdW JJs, it is possible to work in a regime where single CP tunnelling within the JJ is suppressed, whilst second-harmonic Josephson tunnelling becomes instead the dominant contribution to the Josephson energy of the system [178]. As a result, it is possible to realise a qubit that is intrinsically protected, without any requirements for external mechanisms of control or complex engineering [179].

To take advantage of the unique properties offered by superconducting vdW heterostructures in quantum circuits, however, it is also crucial to integrate vdW-based components and devices into more conventional 3D superconducting circuits. This requires, amongst other things, the realisation of transparent and robust superconducting contacts between the vdW and other 3D materials. Sinko and co-workers [180], for example, recently developed a process for the fabrication of electrical contacts with negligible residual resistance between $NbSe_2$ and Al and were able to observe Josephson coupling at the Al/$NbSe_2$ interface. Their process starts with the encapsulation (on both sides) of a $NbSe_2$ flake with hBN in inert ($N_2$) atmosphere, after which the edges of $NbSe_2$ are exposed through a combination of lithography and etching. The sample is then transferred into an evaporation chamber, where first an Ar ion milling is carried out, to clean the cross section of the stack, and then Al is evaporated in-situ (after tilting the sample) to make contact to the edges of $NbSe_2$. The $I_c(B)$ pattern of the JJ investigated in ref. [180] is similar to a Fraunhofer pattern, but with a periodicity that corresponds to an effective area $A_{eff}$ threaded by magnetic flux larger than that calculated based on the device geometry $A_{geom}$. One reason for this discrepancy is that the magnetic penetration depth for ultrathin $NbSe_2$ flakes with thickness $d$ is the Pearl length, which is defined as $\lambda_{Pearl} = 2\lambda^2/d$ [181], other than the London penetration depth $\lambda$. From $\lambda \sim 124$ nm for bulk $NbSe_2$ [182], $\lambda_{Pearl} \sim 2.6$ μm follows for the 12-nm-thick $NbSe_2$ flake studied in ref. [180], meaning that the $NbSe_2$ is uniformly penetrated by $B_{ext}$. In addition, the authors argued that the path along which the superconducting phase winds by $2\pi$ depends on the distribution of the screening current in $NbSe_2$, which in turn depends on the position of the Al leads and on the shape and size of the $NbSe_2$ flake. All these factors can contribute to make $A_{eff}$ much larger than $A_{geom}$. VdW JJs with a large $A_{eff}/A_{geom}$ are advantageous because they combine a small geometric area – this enables high spatial resolution – with the flux sensitivity of a larger JJ. These vdW JJs with high $A_{eff}/A_{geom}$ ratio can therefore find application for the realisation of miniaturised ultrasensitive SQUIDs like those embedded in scanning magnetometers. The as-realised scanning magnetometers would require a single vdW JJ on their scanning tip other than a SQUID, unlike modern scanning SQUID microscopes [183].

The demonstration of Josephson coupling in $NbSe_2$/Al junctions reported in ref. [180] can be also helpful from a fundamental point of view, as a starting point to do in-depth studies aimed at understanding the exact pairing symmetry of $NbSe_2$, which has been suggested to have an odd-parity component of its superconducting order parameter [184,185]. Similar measurements of JJs formed by an *s*-wave S (e.g., Al, Pb, Nb etc.) with a *d*-wave S (e.g., cuprates) have in fact been crucial to demonstrate the *d*-wave symmetry of the order parameter in cuprates [186].

## 11. Outlook and future directions

One of the main advantages of vdW superconducting materials for the realisation of novel devices for superconducting electronics lies in their ability to unlock new physical regimes which are difficult to access with traditional 3D superconducting systems. Their clean, atomically flat interfaces – free of dangling bonds – minimise scattering and decoherence, thus providing an excellent foundation for quantum coherence. Also, the larger upper critical field $B_{c2}$ of vdW Ss down to the monolayer can enable the fabrication of superconducting devices that are very stable against magnetic noise. Thanks to their inherent electrostatic tuneability, superconducting vdW hybrids can also enable gate-controlled devices, reconfigurable JJs, and switchable circuit functionalities — these are key ingredients for the next generation of superconducting logic and quantum computing architectures.

The exceptional flexibility of vdW heterostructures also provides an ideal platform to explore unconventional and topological superconductivity. The ability to engineer precise stacking sequences, where ad-hoc choices can be made in the properties of the vdW materials to couple to a certain vdW S, combined with the possibility of tuning the surface properties of a vdW S, e.g., through molecular adsorption, opens unexplored pathways towards the realisation of spin-triplet pairing or chiral superconductivity. This investigation can lead to devices with many more functionalities and forms of control compared to equivalent ones that can be made with 3D material hybrids (figure 9) and/or help demonstrate clearer evidence for the realisation of exotic states like MZMs that have long been sought for topologically protected qubits and fault-tolerant quantum computing schemes.

From a circuit-related perspective, the high kinetic inductance of vdW Ss like $NbSe_2$ can enable the fabrication of compact low-loss superinductors. These components are highly desirable for fluxonium qubits and high-impedance microwave resonators. Moreover, the possibility of transferring vdW materials in any desired areas paves the way for monolithic vertical integration in superconducting circuits, allowing the replacement of bulky capacitors and tunnel junctions with atomically thin and crystalline elements. Such architectures could lead to superconducting circuits with drastically reduced footprints — an especially compelling prospect for densely-packed logic circuits or qubit arrays, where the minimisation of crosstalk between individual components is critical.

While the integration of vdW materials into superconducting architectures is still in its early stages, it holds great potential. One key milestone for the realisation of vdW-based superconducting logics is the demonstration of vdW-based GCS devices, where the $I_c$ suppression is purely driven by an applied $E$. This is more difficult to achieve in current GCS devices based on gated nanoscale-size superconducting constrictions of 3D Ss, where the GCS seems to originate from a $V_G$-induced $I_{leak}$ in most cases. For vdW-based devices instead, the GCS can be purely driven by an applied $E$, especially because $E$ may be not totally screened in an ultrathin vdW S (e.g., in a monolayer S). Demonstrating such type of GCS devices would lead to GCS logics interfaceable with CMOS and better than current-controlled logics like rapid single flux quantum logics [164,187].

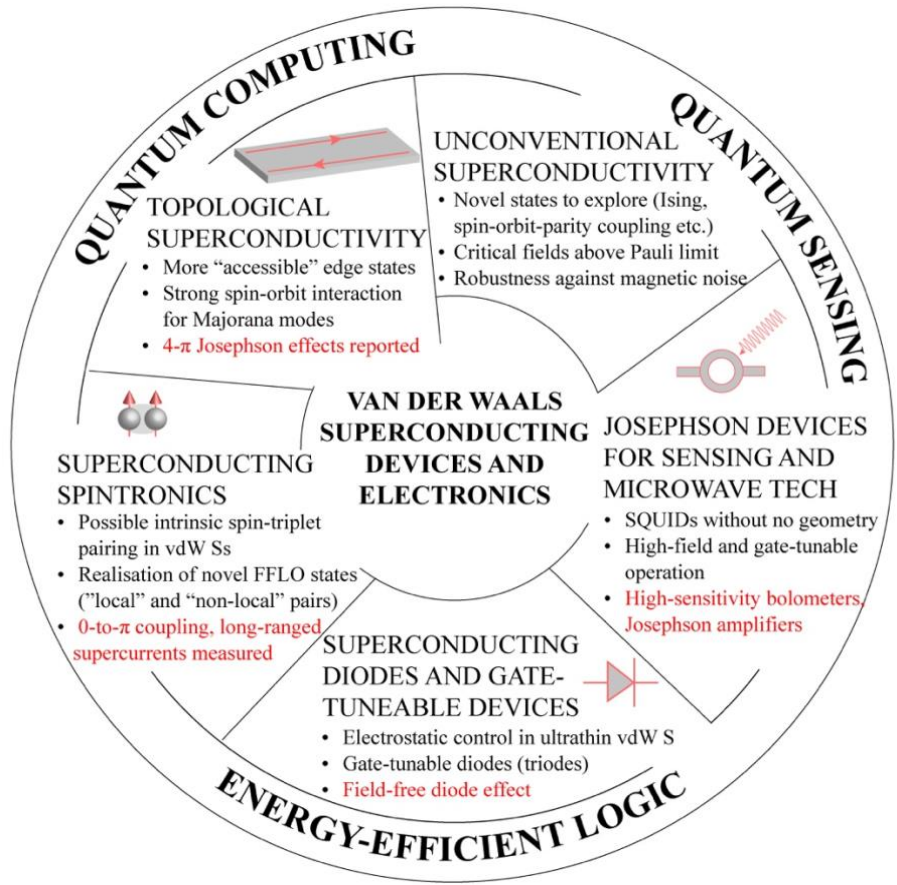

**Figure 9.** Schematic overview of the main physical phenomena and device implementations based on vdW superconductors, together with their corresponding technological domain of application (outer ring). Each sector highlights a representative phenomenon or device enabled by vdW superconductivity, highlighting the specific advantages stemming from vdW superconductors over conventional 3D superconductors. The most significant experimental demonstration reported to date for each area are also listed in red.

Another key milestone for vdW-based quantum circuits is the realisation of vdW-based qubits that combine all functional components including JJs with vdW tunnel barriers (e.g., $MoS_2$ or $WSe_2$), vdW superinductors (e.g., $NbSe_2$-based), and vdW shunt capacitors (e.g., capacitors with hBN or TMD dielectrics). Such devices could drastically reduce qubit footprint and enable new device geometries that are difficult to realise with conventional 3D materials.

In the near term, hybrid systems combining vdW materials with established 3D superconducting platforms would offer a pragmatic pathway to exploit the unique advantages of vdW systems whilst maintaining their compatibility with more mature circuit technologies. However, important challenges remain such as mitigating interfacial disorder and ensuring long-term stability of vdW devices under cryogenic and high-frequency conditions, whilst achieving their fabrication over wafer-scale areas with high uniformity and reproducibility.

Addressing these challenges will require a coordinated effort across materials science, condensed matter physics, and device engineering. Progress in deterministic assembly techniques, encapsulation strategies, and scalable growth methods will be essential to transform vdW superconducting devices from academic prototypes into practical components. Continued exploration of hybrid and all-vdW architectures, along with tailored heterostructure designs, could position vdW-based electronics as a central pillar of future superconducting and quantum technologies.